\documentclass[aps,prd,showpacs,notitlepage,nofootinbib,superscriptaddress,floatfix,showkeys,twocolumn]{revtex4-1}

\usepackage{blindtext}
\usepackage{hyperref}
\usepackage{amsmath,amssymb}
\usepackage{float}
\usepackage{microtype}
\usepackage{ragged2e}  

\usepackage{graphicx}
\usepackage{bm}
\usepackage{latexsym}
\usepackage{epsfig}
\usepackage{psfrag}
\usepackage[dvipsnames]{xcolor}
\usepackage{subfigure}
\usepackage{subcaption}
\usepackage{graphicx}
\usepackage{multirow}
\usepackage{amssymb}
\usepackage{amsmath}
\usepackage{bm}
\usepackage{latexsym}
\usepackage{epsfig}
\usepackage{psfrag}
\usepackage[normalem]{ulem}
\usepackage{textcomp}
\usepackage{color}
\usepackage{pstricks}
\usepackage[utf8]{inputenc}
\usepackage{comment}
\usepackage{hyperref}
\usepackage[mathlines]{lineno}
\usepackage{hyperref}
\usepackage[all]{hypcap}
\hypersetup{  
    colorlinks=true,
    linkcolor=blue,
    filecolor=red,      
    urlcolor=cyan,
    citecolor=green,
    }

\def\beq{\begin{equation}}
\def\eeq{\end{equation}}
\def\bea{\begin{eqnarray}}
\def\eea{\end{eqnarray}}
\def\be{\begin{equation}}
\def\ee{\end{equation}}

\def\bse{\begin{subequations}}
\def\ese{\end{subequations}}

\def\ee{\eta_{\rm e}}

\def\Mpl{M_{_\mathrm{P}}}
\def\f{\frac}
\def\l{\left}
\def\r{\right}
\def\d{\mathrm{d}}

\def\w{w_{\phi}}

\begin{document}
\title{ACT DR6 Insights on $\alpha$-Attractor Inflationary Models and Reheating}

\author{Md Riajul Haque}
\email{riaj1994sjtu.edu.cn}
\affiliation{Tsung-Dao Lee Institute \& School of Physics and Astronomy, Shanghai Jiao Tong University, Shanghai 201210, China}
\affiliation{Physics and Applied Mathematics Unit, Indian Statistical Institute, 203 B.T. Road, Kolkata 700108, India}

\author{Sourav Pal}
\email{soupal1729@gmail.com}
\affiliation{Physics and Applied Mathematics Unit, Indian Statistical Institute, 203 B.T. Road, Kolkata 700108, India}

\author{Debarun Paul}
\email{debarun31paul@gmail.com}
\affiliation{Physics and Applied Mathematics Unit, Indian Statistical Institute, 203 B.T. Road, Kolkata 700108, India}

\begin{abstract}
We investigate the observational constraints on $\alpha$-attractor inflationary models and their post-inflationary reheating dynamics in light of the latest CMB data from ACT DR6 combined with Planck18, BICEP/$Keck$ 2018, and DESI (collectively denoted P-ACT-LB-BK18). Focusing on both E- and T-type attractor potentials, we analyze how inflationary observables, namely the scalar spectral index $n_s$ and the tensor-to-scalar ratio $r$, are indirectly influenced by reheating parameters such as the reheating temperature $T_{\text{RH}}$, the inflaton equation of state $w_\phi$, and consequently the inflaton’s couplings to Standard Model (SM) particles. We incorporate constraints from primordial gravitational wave (PGW) overproduction via $\Delta N_{\rm eff}$ bounds, which are particularly relevant for stiff post-inflationary dynamics. Additionally, we include theoretical constraints from Coleman--Weinberg (CW) one-loop radiative corrections, which impose an upper bound on the couplings, as well as inflaton self-resonance, a non-perturbative effect that can occur even in the absence of inflaton--SM interactions and imposes a strong lower limit, particularly significant for $1/3 \lesssim w_\phi \lesssim 0.7$. Our analysis shows that E-models allow a broad range of reheating scenarios, including matter-like reheating ($w_\phi = 0$), whereas the T-model yields more restrictive results and remains viable only for $w_\phi > 1/2$ for all three non-gravitational interaction channels, $\phi \to \bar{f}f$, $\phi \to bb$, and $\phi\phi \to bb$ are considered. For stiffer equations of state, an intermediate coupling window survives, bounded from above by CW corrections or $2\sigma$ observational limits, and from below by the combined PGW and self-resonance constraints. We derive updated bounds on inflaton couplings for both decay and scattering channels and identify parameter regions consistent with successful perturbative reheating. These results establish a robust connection between inflationary dynamics, reheating physics, and particle interactions, and provide concrete targets for upcoming precision CMB observations.
\end{abstract}

\maketitle
\tableofcontents

\black

\section{Introduction}
\label{sec:intro}
 
The last decade has witnessed an unprecedented leap in our understanding of the early Universe, primarily driven by high-precision measurements of the cosmic microwave background (CMB)~\cite{CMB-S4:2016ple,Planck:2018jri} . Observations from Planck18, BICEP/$Keck$ 2018, and more recently, the Atacama Cosmology Telescope (ACT), have dramatically tightened  constraints on the inflationary models. These advances not only refine our estimates of key inflationary observables, such as the scalar spectral index $n_s$ and the tensor-to-scalar ratio $r$, but also enable us to probe the microphysics underlying inflation~\cite{Guth:1982ec,Lazarides:1997xr,Starobinsky:1982ee} and its aftermath, particularly the reheating epoch~\cite{Kolb:1990vq,Shtanov:1994ce,Kofman:1997yn,Allahverdi:2010xz,Lozanov:2019jxc}. 

Among the diverse landscape of inflationary frameworks, $\alpha$-attractor models~\cite{Kallosh:2013yoa,Kallosh:2013hoa,Roest:2013fha,Ferrara:2013rsa,Kallosh:2013tua,Cecotti:2014ipa,Kallosh:2014rga,Galante:2014ifa} have received considerable attention due to their strong theoretical underpinnings and robust predictive power. These models arise naturally in the contexts of supergravity and string theory, where the parameter $\alpha$ encapsulates the curvature of the scalar field manifold. By smoothly interpolating between chaotic inflation~\cite{Starobinsky:1980te,Linde:1983gd} and plateau-like scenarios, such as Starobinsky inflation~\cite{Ellis:2013nxa,Kallosh:2014laa}, $\alpha$-attractors offer a unifying structure that encompasses a wide range of potentials. This makes them a compelling framework for connecting fundamental theory with cosmological observations.

The recent Data Release 6 (DR6) from the Atacama Cosmology Telescope (ACT) has provided improved measurements of the high-$\ell$ CMB power spectrum. When combined with Planck data (P-ACT), the scalar spectral index is constrained to $n_s = 0.9709 \pm 0.0038$~\cite{ACT:2025fju,ACT:2025tim}. Further inclusion of CMB lensing and Baryon Acoustic Oscillation (BAO) data from DESI (P-ACT-LB) refines the constraint to $n_s = 0.9743 \pm 0.0034$ \cite{ACT:2025fju,ACT:2025tim}, which deviates from the Planck-only result at the $2\sigma$ level. This mild tension disfavors classic plateau models such as Starobinsky inflation, while favoring scenarios that predict slightly higher values of $n_s$, including $\alpha$-attractor models. These updated constraints necessitate a renewed evaluation of the inflationary landscape, particularly for models that previously lay near the boundary of observational viability. Motivated by these developments, a variety of inflationary scenarios, such as warm inflation, Higgs inflation, and other theoretically motivated frameworks, have been revisited in the context of the recent ACT data~\cite{Kallosh:2025rni,Aoki:2025wld,Berera:2025vsu,Dioguardi:2025vci,Gialamas:2025kef,Salvio:2025izr,Antoniadis:2025pfa,Haque:2025uis,Haque:2025uga,Kim:2025dyi,Dioguardi:2025mpp,Gao:2025onc,He:2025bli,Pallis:2025epn,Drees:2025ngb}.

However, the inflationary observables $n_s$ and $r$ alone are insufficient to fully reconstruct the dynamics of the early Universe. The reheating phase, during which the inflaton decays and the Universe transitions into a hot, radiation-dominated era, plays a crucial intermediary role. It sets the initial conditions for Big Bang Nucleosynthesis (BBN), influences the number of inflationary e-folds between horizon exit of a given CMB pivot scale and the end of inflation, and affects the predicted values of $n_s$ and $r$. Moreover, theoretical models of reheating establish a vital connection between inflation and particle physics via inflaton couplings to Standard Model (SM) or beyond the Standard Model (BSM) fields. This phase is typically characterized by two key parameters: the reheating temperature $(T_{\rm RH})$ and the effective equation-of-state (EoS) $(w_\phi)$. The reheating temperature depends sensitively on both the inflaton’s coupling to SM and the structure of the inflaton potential $V(\phi)$ near its minimum—often determined by the exponent in the potential.

In this work, we consider reheating scenarios in which the inflaton decays into either fermions ($\phi \to \bar{f}f$) or bosons ($\phi \to bb$), as well as scattering processes such as $\phi\phi \to bb$ \cite{Garcia:2020wiy,Garcia:2020eof,Haque:2023yra}. The decay products are assumed to be massless and rapidly thermalize, contributing to the radiation bath. Due to the absence of direct observational probes of this epoch, the reheating temperature $T_{\rm RH}$ remains poorly constrained. It is bounded from above by the energy scale allowed by CMB observations, corresponding to instantaneous reheating ($\sim 10^{15}$ GeV), and from below by the minimum temperature required for successful BBN ($T_{\rm BBN} \sim 4$ MeV)~\footnote{Incorporating galaxy survey data with BBN and CMB observations modifies the lower constraint on the reheating temperature, as indicated in~\cite{Barbieri:2025moq,deSalas:2015glj}.}~\cite{Kawasaki:1999na,Kawasaki:2000en,Hasegawa:2019jsa}. While the standard inflationary observables ($n_s,\,r$) are shaped by both inflationary and post-inflationary dynamics, the details of reheating encode complementary information—particularly about fundamental couplings between the inflaton and SM particles \cite{Garcia:2020wiy,Garcia:2020eof,Haque:2023yra,Giudice:2000ex,Maity:2018dgy,Haque:2019prw,Haque:2020zco,Mambrini:2021zpp,Haque:2021mab,Clery:2021bwz,Haque:2022kez}.

Recently, there have been growing efforts to constrain reheating parameters using BICEP/$Keck$ 2018 (BK18) data in combination with Planck18 observations \cite{Ellis:2021kad,Drewes:2023bbs,Chakraborty:2023ocr}. In this paper, we explore constraints on both the $\alpha$-attractor model parameters, reheating parameters—including $T_{\rm RH}$, the inflaton EoS $w_\phi$, and inflaton couplings—using the latest ACT DR6 data in combination with Planck18, BK18, and DESI. Additionally, we incorporate an independent constraint arising from the overproduction of primordial gravitational waves (PGWs). The PGWs correspond to tensor perturbations generated due to vacuum fluctuations during inflation~\cite{Grishchuk:1974ny,Starobinsky:1979ty,Boyle:2005se,Watanabe:2006qe,Saikawa:2018rcs,Caprini:2018mtu}. The dynamics of reheating leave imprints on the stochastic background of these PGWs, particularly in modes that re-enter the horizon during this phase~\cite{Bernal:2019lpc,Haque:2021dha,Maity:2024odg,Maity:2024cpq,Ghoshal:2024gai,Barman:2023ktz}. A stiff EoS ($w_\phi \geq 1/3$) leads to a blue-tilted GW spectrum, potentially resulting in an excess of high-frequency GWs. These excess PGWs contribute to the effective number of relativistic degrees of freedom, $\Delta N_{\rm eff}$, which is constrained by both BBN and CMB data. Consequently, $\Delta N_{\rm eff}$ bounds \footnote{For a comprehensive summary of current and predicted constraints on the effective number of relativistic species, $\Delta N_{\rm eff}$, we refer the reader to Table II of Ref.~\cite{Barman:2024slw}. In this analysis, however, we adopt the P-ACT-LB bound, which imposes a constraint of $\Delta N_{\rm eff} < 0.17$ \cite{ACT:2025fju,ACT:2025tim} at 95\% C.L.} offer an indirect probe of reheating physics, allowing us to place a lower bound on the reheating temperature $T_{\rm RH}$ for stiff post-inflationary dynamics.

Beyond these observational bounds, theoretical effects further constrain the viable parameter space. In particular, Coleman--Weinberg (CW) one-loop radiative corrections to the inflaton potential~\cite{Coleman:1973jx,Armendariz-Picon:2020tkc,Bastero-Gil:2016qru,Gorgulho:2025wxz} play a crucial role at large coupling values, where they can significantly distort the effective potential from its tree-level form~\cite{Kallosh:2013tua,Kallosh:2013yoa,Wolf:2025ecy,Gialamas:2025kef}. As a result, the standard predictions obtained from the tree-level potential are no longer reliable in this regime. Requiring theoretical consistency of the inflationary potential therefore imposes a stringent upper bound on the coupling parameters \cite{Chakraborty:2023ocr}. On the other hand, non-perturbative dynamics such as inflaton self-resonance~\cite{Amin:2010dc,Amin:2010xe,Amin:2011hj,Lozanov:2017hjm} can also play an important role in the post-inflationary epoch. In this process, small perturbations in the homogeneous inflaton field grow exponentially, leading to unstable oscillations and efficient particle production through non-linear effects. Notably, self-resonance can occur even in the absence of any direct coupling between the inflaton and SM fields, implying that the conventional perturbative reheating framework may break down in this regime. Consequently, requiring reheating to proceed dominantly through perturbative processes imposes a complementary lower bound on the coupling parameters. Within the parameter space favored by our combined observational and theoretical constraints, reheating predominantly remains in the perturbative regime, however, the self-resonance condition provides the most stringent lower bound, particularly for $1/3 \lesssim w_\phi \lesssim 0.7$.
 
This work is motivated by the synergy between recent ACT DR6 measurements and theoretical developments in inflationary reheating. Our main objectives are as follows:
\begin{itemize}
    \item To analyze the viability of $\alpha$-attractor models—both $E$- and $T$ type in light of updated constraints on $n_s$ and $r$ from the combination of Planck18, ACT, DESI, and BICEP/$Keck$ 2018 data.
    
    \item To explore the implications of various reheating scenarios, parametrized by different EoS and inflaton decay channels, on the inflationary parameter space.
    
    \item To incorporate indirect constraints on the reheating temperature arising from the overproduction of PGWs, particularly through bounds on the effective number of relativistic species, $\Delta N_{\rm eff}$, along with additional restrictions from self-resonance.
    
    \item To derive bounds on inflaton couplings to Standard Model particles through bosonic and fermionic channels, which determine the efficiency of reheating and shape the subsequent thermal history of the Universe. In doing so, we account for the fact that in certain cases the upper limits on the couplings are set by Coleman-Weinberg one-loop radiative corrections, beyond which the analysis based on the tree-level potential is no longer valid. On the other hand, the lower limits are constrained by consistency with the BBN bound, PGW overproduction, and restrictions from self-resonance.
\end{itemize} 

The structure of the paper is as follows. In Sec.~\ref{sec:inflation_overview}, we provide an overview of the inflationary framework, focusing on the $\alpha$-attractor models. Section~\ref{sec:reheating} presents a detailed analysis of the reheating phase, including inflaton decay via fermionic and bosonic channels. In Sec.~\ref{sec:delta_neff}, we investigate constraints on the reheating temperature and the inflaton EoS arising from the overproduction of PGWs. We then turn to theoretical bounds: Sec.~\ref{sec:loop_correction} discusses one-loop radiative corrections to the inflationary potential, while Sec.~\ref{sec:self_resonance} explores the regime where self-resonance dominates and the perturbative reheating framework is no longer valid. Section~\ref{sec:result} presents our complete results, including constraints on the inflaton potential parameters and the associated reheating quantities, $T_{\rm RH}$ and $w_\phi$, based on ACT DR6 data, incorporating limits from PGWs and non-perturbative effects, as well as identifying the ranges of inflaton--SM coupling parameters consistent with these constraints. Finally, Sec.~\ref{sec:conclusion} summarizes our findings and outlines directions for future research.

\section{Overview of the inflationary model}
\label{sec:inflation_overview}

In the context of canonical single-field inflationary theories, two well-studied frameworks, the T-model and the E-model describe the dynamics of the inflaton through attractor-type potentials. These models are characterized by the scalar potentials,
\begin{equation}
\label{eq:attractorpotential}
V(\phi)  \;=\; 
\begin{cases}
\Lambda^4\,\left[1-e^{-\sqrt{\frac{2}{3\,\alpha}}\phi/M_{\rm P}}\right]^{2 n}\,, & {\rm E - model}\,,\\[10pt]
\Lambda^4\,\left[\tanh\left(\frac{\phi}{\sqrt{6\alpha}\,M_{\rm P}}\right)\right]^{2n}\,,&  {\rm T-model}\,. \\[10pt]
\end{cases}
\end{equation}
Here, $\Lambda$ denotes the inflationary energy scale, $n$ is index of the potential controlling the steepness near its minimum, and $\alpha$ is a dimensionless parameter shaping the curvature of the potential \footnote{Note that $M_{\rm P}$ is the reduced Planck mass, defined as $ M_{\rm P} \equiv \frac{1}{\sqrt{8\pi G}} = 2.43 \times 10^{18} \, \text{GeV} $.}.
These model parameters can be essentially determined from the scalar curvature perturbations generated during inflation, characterized by the power spectrum $\Delta^2_{\mathcal{R}}  =A_{\mathcal{R}} (k/k_{\ast})^{n_s -1}$. Here, $A_{\mathcal{R}}$ is the amplitude of scalar fluctuations and $n_s$ is the scalar spectral index. $k_*$ is the pivot scale which is considered to be $0.05 \, {\rm Mpc}^{-1}$ throughout our analysis. Another key CMB observable is the tensor-to-scalar ratio, defined as $r \equiv A_{\mathcal{T}} / A_{\mathcal{R}}$, where $A_{\mathcal{T}}$ denotes the amplitude of the tensor power spectrum. The tensor spectrum is typically nearly scale-invariant, arising from quantum fluctuations of the gravitational field during inflation.
For any canonical single field inflationary model, the slow-roll parameters can be defined as,
\begin{equation}\label{eq:epeta}
\epsilon \equiv \frac{1}{2} M_{\rm p}^{2}\left(\frac{V^{\prime}}{V}\right)^{2} \, , \qquad \eta \equiv M_{\rm p}^{2}\left(\frac{V^{\prime \prime}}{V}\right) \, ,
\end{equation}
where prime denotes derivative with respect to $\phi$. Furthermore, under the slow-roll approximation, CMB observables ($n_s,r$ and $A_s$) can be expressed in terms of the slow-roll parameters as follows,
\begin{equation}
n_{s}= 1- 6 \epsilon(\phi)+ 2 \eta(\phi)~,~r=16\epsilon(\phi)\,.    
\end{equation}
Subsequent dynamics of the inflaton field is governed by the inflationary energy scale /Hubble parameter ($H_k$), which can be approximated in terms or $r$ and $A_{\mathcal{R}}$ as follows,
\begin{eqnarray}
H_{\rm k} &=& \frac{\pi M_{\rm P}\sqrt{r\,A_{\mathcal R}}} {\sqrt{2}} \simeq  \sqrt{\frac{V(\phi_{\rm k})}{3 M_{\rm P}^2} } \, .\label{eq:Hk} 
\end{eqnarray} 
The duration of inflation can be parametrized by the total e-folds number ($N_k$) between the horizon crossing of the CMB mode with comoving number $k$ and the end of inflation. For the $\alpha$-attractor model, $N_k$ is given by \cite{Drewes:2017fmn,Garcia:2020wiy,German:2021tqs},
\begin{equation}
\label{eq:Nk}
N_k =
\begin{cases}
\frac{3 \alpha}{4 n}\left(\exp{\sqrt{\frac{2}{3\alpha}}\frac{\phi_{k}}{M_{\rm P}}}-\exp{\sqrt{\frac{2}{3\alpha}}\frac{\phi_{\rm end}}{M_{\rm P}}} \right) \\ 
\quad  -\sqrt{\frac{3\alpha}{8 n^2 M_{\rm P}^2}}(\phi_k - \phi_{\rm end})\,,   & {\rm E - model}\,, \\
\frac{3\alpha}{4n}\left(\cosh{\sqrt{\frac{2}{3\alpha}}\frac{\phi_k}{M_{\rm P}}}-\cosh{\sqrt{\frac{2}{3\alpha}}\frac{\phi_{\rm end}}{M_{\rm P}}} \right)\, & {\rm T - model}\,,
\end{cases}
\end{equation}
where $\phi_k$ and $\phi_{\rm end}$ are field values at the time of horizon crossing and at the end of inflation, respectively. One can obtain the value of $\Lambda$ in the large field limit by setting $\Lambda^4 \approx V(\phi_k)$. 
More general expressions for $\phi_k$ and $\Lambda$ can be found in \cite{Drewes:2017fmn} for the E-model, and in \cite{Garcia:2020wiy} for the T-model. The end of the inflation is marked by the condition $\epsilon(\phi_{\rm end})=1$.
Finally the tensor-to-scalar ratio, $r$ can be expressed in terms of the $\alpha$-attractor model parameters as follows \cite{Drewes:2017fmn,Garcia:2020wiy,German:2021tqs},
\begin{equation}
r\;=\;
\begin{cases}\label{eq:r_model}
\frac{64n^2}{3\alpha}\left(e^{\sqrt{\frac{2}{3\,\alpha}}\frac{\phi_k}{M_{\rm P}}}-1\right)^{-2}\,,& {\rm E- model}, \\
\frac{64n^2}{3\alpha} \text{csch}^2\left(\sqrt{\frac{2}{3\,\alpha}}\frac{ \phi _k}{M_{\rm P}}\right) \,,& {\rm T- model}. 
\end{cases}
\end{equation}
After inflation, the Universe is cold, dark, and dominated by the homogeneous inflaton field. A natural consequence of this state is the decay of the inflaton into SM particles, thereby reheating the Universe and initiating a radiation-dominated era, an essential condition for the onset of BBN. The inflaton energy density at the end of inflation can be approximated as~\cite{Chakraborty:2023ocr,Garcia:2020wiy},
\begin{equation}\label{eq:ic}
\rho^{\rm end}_{\phi} 
\sim \frac{\Lambda^4}{\alpha_1^{2n}} \left[\frac{\phi_{\rm end}}{M_{\rm P}}\right]^{2 n}  ,
\end{equation}
where $\alpha_1 \equiv (\sqrt{3\alpha /2},\,\sqrt{6\alpha})$ for E-model and T-model, respectively. This relation allows us to connect the inflationary era with the subsequent evolution during the post-inflationary epoch, namely the reheating phase. Let us now briefly discuss the reheating phase, highlighting how various non-gravitational couplings influence both the duration of reheating and the temperature at its conclusion, commonly referred to as the reheating temperature $T_{\rm RH}$.

\section{Overview of Reheating Dynamics and Possible Scenarios}
\label{sec:reheating}
Following the end of inflation, the inflaton field undergoes coherent oscillations around the minimum of its potential. The potential near this minimum behaves as $V(\phi) \sim \phi^{2n}$ for the $\alpha$-attractor models considered here. These oscillations mark the beginning of the reheating era, where the inflaton decays into SM particles, gradually transitioning the Universe into a radiation-dominated (RD) phase.

The inflaton field that oscillates coherently, may be broken down into
\begin{equation}
\label{eq:inflaton_decompose}
\phi(t) = \phi_0(t) \mathcal{P}(t),
\end{equation}
where $\phi_0(t)$ denotes the slowly varying envelope, incorporating the effects of redshifts, and $\mathcal{P}(t)$, a quasi-periodic function, describes the short timescale oscillation of the potential with Fourier expansion $\mathcal{P}(t) = \sum_{\mu=-\infty}^{\infty} \mathcal{P}_\mu e^{i\mu \omega t}$. The oscillation frequency is given by~\cite{Garcia:2020wiy}
\begin{equation}
\label{eq:freq_inflaton_osc}
\omega = m_\phi(t) \sqrt{\frac{\pi n}{(2n-1)}} \frac{\Gamma\left(\frac{1}{2} + \frac{1}{2n}\right)}{\Gamma\left(\frac{1}{2n}\right)},
\end{equation}
where $m_{\phi}$ is the mass of the inflaton, which can defined as,
\begin{eqnarray}
\label{eq:m_phi}
    m_\phi(t)\equiv\sqrt{V''(\phi_0(t))}\propto \phi_0(t)^{2(n-1)}.
\end{eqnarray}

Assuming that the oscillation period is much shorter than the timescales associated with redshift and decay, averaging over a single oscillation yields an inflaton energy density approximately given by $\rho_\phi \sim V(\phi_0)$, and determines the effective average EoS ~\cite{Ford:1986sy, Garcia:2020wiy}
\begin{equation}
\label{eq:EoS}
\w \simeq \frac{n - 1}{n + 1}.
\end{equation}

Considering that the inflaton decays only into the radiation, the Boltzmann equation for the energy density of inflaton ($\rho_{\phi}$) and radiation ($\rho_R$) can be expressed as
\begin{align}
\dot{\rho}_\phi + 3H(1 + w_\phi) \rho_\phi &= -\Gamma_\phi(t) (1 + w_\phi) \rho_\phi, \label{eq:boltzmann_eq1}\\
\dot{\rho}_R + 4H \rho_R &= \Gamma_\phi(t) (1 + w_\phi) \rho_\phi,\label{eq:boltzmann_eq2}
\end{align}
with $\Gamma_{\phi} (t)$ being the time dependent decay rate of inflaton. The corresponding Hubble rate can be determined as follows:
\begin{equation}
\label{eq:hubble}
H^2 = \frac{\rho_\phi + \rho_R}{3 \Mpl^2}.
\end{equation}
Reheating phase ends when $\rho_{\phi}\simeq\rho_R$, 
where the equality marks the temperature of reheating ($T_{\rm RH}$) with corresponding scale factor $a_{\rm RH}$. For analytical estimation, the Boltzmann equations (Eqs.~\eqref{eq:boltzmann_eq1}) yield $\rho_\phi \sim a^{-3(1+w_{\phi})}$, which governs the background dynamics and is the main determining factor for the Hubble parameter and  the radiation energy density follows $\rho_{\rm R} \propto T^4$. Thus, the radiation temperature at the end of reheating \textit{i.e.}, the reheating temperature, can be expressed as
\begin{eqnarray}
\label{eq:trh_w_nrh}
    T_{\rm RH} \simeq \left(\frac{90 \Mpl^2 H_{\rm end}^2}{\pi^2g_{\ast RH}}\right)^{1/4} e^{-\frac{3}{4}N_{\rm RH}(1+\w)},
\end{eqnarray}
Here, $N_{\rm RH}$ denotes the number of e-folds between the end of inflation and the end of the reheating phase. The case $N_{\rm RH} = 0$ corresponds to an instantaneous reheating scenario. The quantity $g_{\ast \rm RH}$ represents the effective number of relativistic degrees of freedom in the thermal bath at the end of reheating.
Under the standard assumption, the comoving entropy density of the Universe is conserved from the end of reheating until the present-day. Thus, assuming the conservation of comoving entropy density, the reheating temperature can be connected with the present-day temperature of the Universe $T_0$ ($=2.725$ K) as ~\cite{Dai:2014jja,Cook:2015vqa}
\begin{equation}
\label{eq:trh_w_nrh_ne}
T_{\rm RH} = \left(\frac{43}{11 g_{*S,\rm RH}}\right)^{1/3} T_0 \frac{H_{\rm k}}{k_{\ast}} e^{-(N_{k} + N_{\rm RH})},
\end{equation}
where $N_{k}$ denotes the number of e-folds between the end of inflation and the horizon exit of a given CMB pivot scale, $k_{\ast}$. $g_{*S,\rm RH}$ is the relativistic degree for freedom associated with the entropy calculated at the end of reheating. Upon comparison of Eq.~\eqref{eq:trh_w_nrh} and Eq.~\eqref{eq:trh_w_nrh_ne}, $N_{\rm k}$ can be written in terms of reheating temperature as
\begin{align}
\label{eq:Ninf_reheating}
    N_{k} = \,&\log \left[
\left(\frac{43}{11g_{\ast S, RH}}\right)^{1/3} T_0 \frac{H_{\rm k}}{k_{\ast}}  \right . \nonumber \\
& \times \left. T_{\rm RH}^{\frac{4}{3(1 + \w)} - 1} 
\left(\frac{\pi^2g_{\ast RH}}{90 \Mpl^2 H_{\rm end}^2}\right)^{\frac{1}{3(1 + \w)}}
\right]
\end{align}

\begin{figure}[!ht]
    \centering
    \includegraphics[scale=0.8]{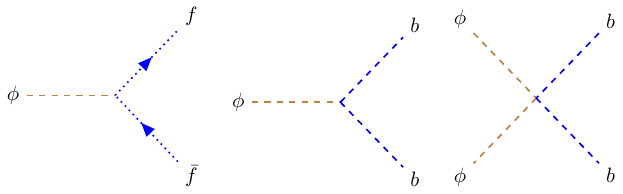}
    \raggedright
    \caption[]{\justifying \it Feynman diagrams for the decay of inflaton into fermions and bosons, and annihilation into bosons.}
    \label{fig:feynman}
\end{figure}
Depending on how the inflaton couples to other fields, reheating can proceed via non-gravitational interactions. These interactions are governed by specific decay or annihilation processes involving the inflaton, which is illustrated by the Feynman diagram in Fig.~\ref{fig:feynman}. The corresponding interaction Lagrangian for such non-gravitational interactions can be written as~\cite{Garcia:2020wiy}
\begin{eqnarray}
\label{eq:Lint}
    \mathcal{L}_{\rm int} \supset  
    \begin{cases} 
    y_\phi\phi \bar{f}f & \phi \to \bar{f}f \\
    g_\phi \phi bb & \phi \to b b\\
    \sigma_\phi \phi^2 b^2 & \phi \phi\to b b\\
\end{cases},
\end{eqnarray}
with $f$ and $b$ being the fermionic and basonic fields, respectively, however, this can be extended for more inflaton-matter/radiation coupling in a very straightforward fashion. The coupling $y_\phi$ and $\sigma_\phi$ are the dimensionless Yukawa-like coupling and four-point coupling, respectively, whereas, $g_\phi$ is the dimensionful bosonic coupling. The corresponding decay/annihilation rate of the aforementioned interactions can be expressed as~\cite{Garcia:2020wiy}
\begin{eqnarray}
\label{eq:int_rates}
\Gamma_{\phi} (t)\equiv 
\begin{cases}
\frac{y_{\rm eff}^2\,m_\phi(t)}{8\pi}\,, & \phi\rightarrow \bar{f}f\,,\\
\frac{g_{\rm eff}^2}{8\pi\,m_\phi(t)}\,,& \phi\rightarrow bb\,,\\
\frac{\sigma_{\rm eff}^2}{8\pi}\,\frac{\rho_\phi(t)}{m_\phi^3(t)}\,,& \phi\phi\rightarrow bb\,.
\end{cases}
\end{eqnarray}
The `\textit{eff}' at the suffix represents the effective coupling over oscillation average. The effective coupling defers from standard coupling because for $n>1$, the inflaton mass becomes time-dependent due to the oscillatory behavior of the field $\phi(t)$ (see Eq.~\eqref{eq:m_phi}). The ratio of the oscillation-induced effective coupling parameters to their corresponding Lagrangian values can be estimated as~\cite{Garcia:2020wiy}
\begin{align}
\label{eq:yeff_y}
\left(\frac{y_{\rm eff}}{y_\phi}\right)^2&=2(n+1)(2n-1)\,\left(\frac{\omega}{m_\phi}\right)^3\sum^{\infty}_{\mu=1}\mu^3\,\lvert\mathcal P_\mu\rvert^2\,,\\
\left(\frac{g_{\rm eff}}{g_\phi}\right)^2&=2(n+1)(2n-1)\,\frac{\omega}{m_\phi}\sum^{\infty}_{\mu=1}\mu\,\lvert\mathcal P_\mu\rvert^2\,,\\
\left(\frac{\sigma_{\rm eff}}{\sigma_\phi}\right)^2&=4n\,(n+1)\,(2n-1)^2\,\frac{\omega}{m_\phi}\sum^{\infty}_{\mu=1}\mu\,\lvert\mathcal (\mathcal P^2)_\mu\rvert^2\,.
\end{align}
The authors of \cite{Maity:2024cpq} provides the numerical values for the effective coupling for different EoS with corresponding Fourier sum, which is displayed in Table~\ref{tab:eff_coupling}.

\begin{table}[!ht]
\centering
\renewcommand{\arraystretch}{1.2}
 \begin{tabular}{c | c |c |c |c|c |c } 
 \hline
 \hline
 $n\,(w_\phi)$ & $\sum \mu^3\lvert \mathcal{P}_\mu\rvert^2$ & $\sum \mu\lvert\mathcal P_\mu\rvert^2$ & $\sum \mu\lvert(\mathcal P^2)_\mu\rvert^2$  & $\frac{y_{\rm eff}}{y_\phi}$& $\frac{g_{\rm eff}}{g_\phi}$& $\frac{\sigma_{\rm eff}}{\sigma_\phi}$\\ [1.0ex] 
 \hline
 $1 (0)$ & $\frac{1}{4}$ & $\frac{1}{4}$ & $\frac{1}{8}$ & 1 & 1 & 1\\ 
 $2 \left(\frac{1}{3}\right)$ & 0.241  & 0.229& 0.125 & 0.71 & 1.42& 3.64\\
 $5 \left(\frac{2}{3}\right)$& 0.257 & 0.205 & 0.120& 0.50 & 2.14& 15.6 \\
 $7 \left(\frac{3}{4}\right)$ & 0.270 & 0.198 & 0.117&0.44 & 2.49& 25.8\\
 $10 \left(\frac{9}{11}\right)$ & 0.287  & 0.191& 0.114&  0.38 & 2.92& 44.0 \\ 
 \hline
 \hline
 \end{tabular}
 \caption[]{\justifying \it Numerical values for the effective couplings and the Fourier sums.}
 \label{tab:eff_coupling}
\end{table}

\begin{figure*}
\includegraphics[width=0017.50cm,height=013.0cm]{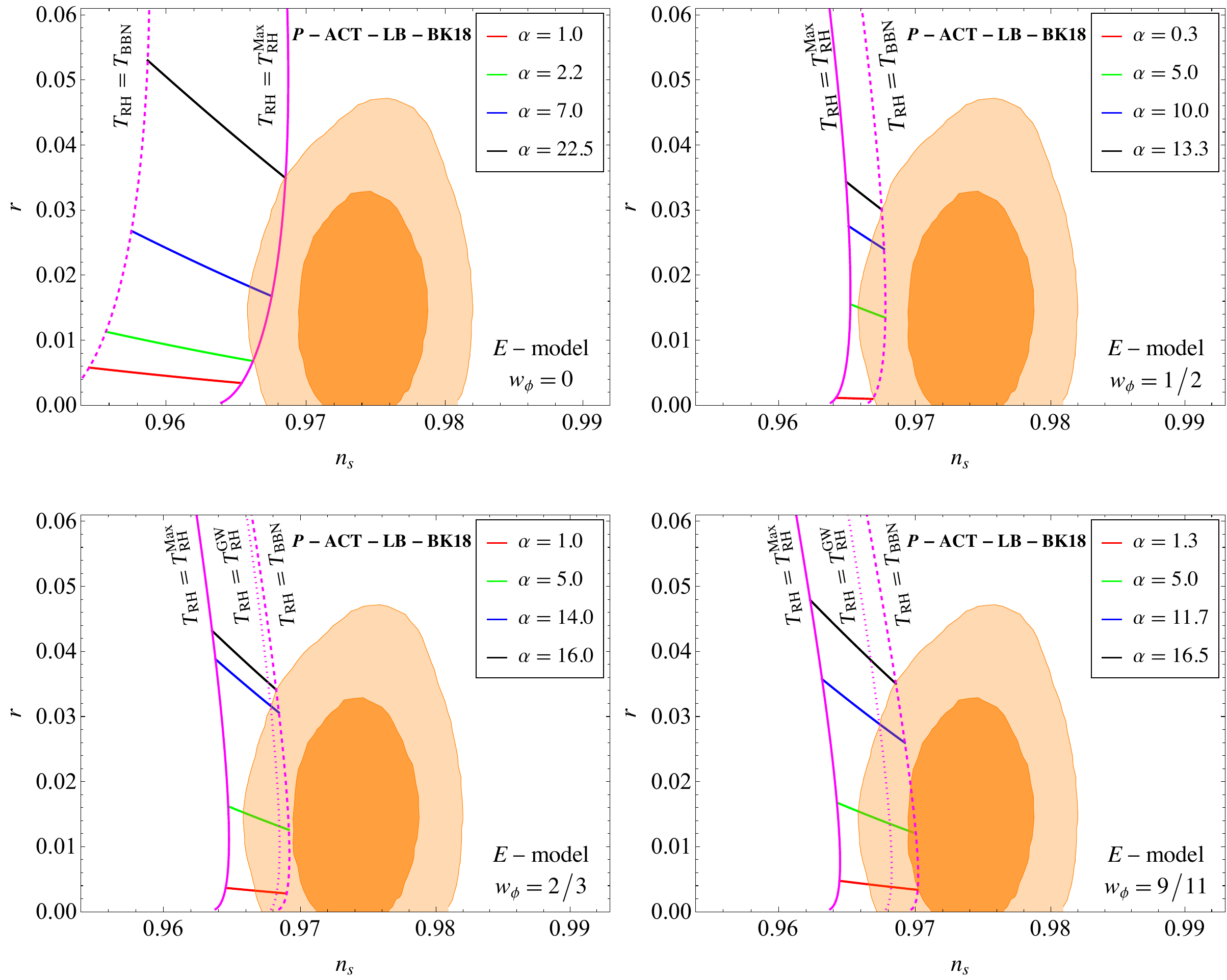}
\caption[]{\justifying \em Predictions of the $\alpha$-attractor \textbf{E-model} for various combinations of $(\alpha,,w_\phi)$ are shown projected onto the $(n_s, r)$ plane, overlaid with the latest combined constraints from {\rm P-ACT-LB-BK18}. Both $n_s$ and $r$ are projected for $k_{\ast}=0.05\,{\rm Mpc}^{-1}$. The deep and light orange shaded regions indicate the $1\sigma$ (68\% C.L.) and $2\sigma$ (95\% C.L.) confidence intervals, respectively. The reheating temperature spans from $T_{\rm BBN}$ to $T_{\rm RH}^{\rm Max}$, represented by dashed and solid magenta lines. An additional critical temperature scale, $T_{\rm RH}^{\rm GW}$, is marked by a dotted magenta line, which is derived from $\Delta N_{\rm eff}$ constraints on the overproduction of PGWs.}
\label{fig:nsrE}
\end{figure*}
\begin{figure*} 
\includegraphics[width=0017.50cm,height=013.0cm]
          {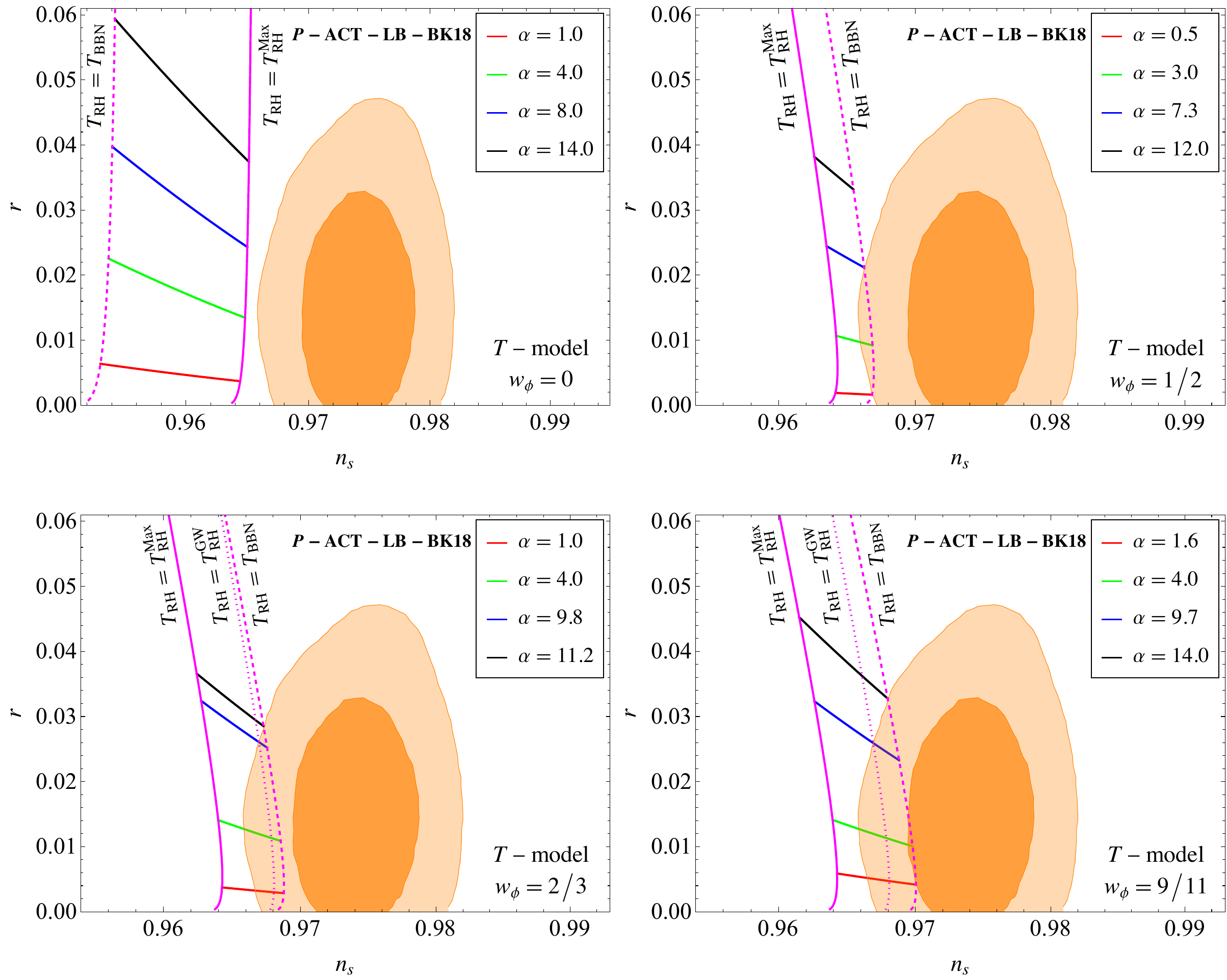}
   \caption[]{\centering \em The description of the figure is same as Fig.~\ref{fig:nsrE} but for \textbf{T-model}.}
\label{fig:nsrT}
\end{figure*}

Using the expression of $m_{\phi}$ (Eq.~\eqref{eq:m_phi}), one can rewritten Eq.~\eqref{eq:int_rates} as follows~\cite{Garcia:2020wiy}:
\begin{eqnarray}
\label{eq:Gammafinal}
\Gamma_\phi(t)\equiv\gamma_\phi(t)\left(\frac{\rho_\phi}{\Mpl ^4}\right)^l\,,
\end{eqnarray}
with
\begin{eqnarray}
\label{eq:gamma}
\gamma_{\phi} (t),\,l \equiv
\begin{cases}
\frac{y_{\rm eff}^2\,\sqrt{2n\,(2n-1)}\,\lambda^{1/2n}\,\Mpl }{8\pi},~~\,\frac{1}{2}-\frac{1}{2n}\, & \text{for }\phi\rightarrow \bar{f}f\,,\\
\frac{g_{\rm eff}^2}{8\pi\,\sqrt{2n\,(2n-1)}\,\lambda^{1/2n}\,\Mpl },\,~~~\frac{1}{2n}-\frac{1}{2}\, & \text{for }\phi\rightarrow bb\,,\\
\frac{\sigma_{\rm eff}^2\,\Mpl }{8\pi\,\left(2n\,(2n-1)\right)^{3/2}\,\lambda^{3/2n}}\,,~~~~~\frac{3}{2n}-\frac{1}{2}\,& \text{for } \phi\phi\rightarrow bb\,.
\end{cases}
\end{eqnarray}
where $\lambda\equiv (2/(3\alpha))^n(\Lambda/M_{\rm P})^4$. Thus, in terms of the interaction rates, the reheating temperature can be estimated as~\cite{Garcia:2020wiy}
\begin{widetext}
\begin{eqnarray}
\label{eq:trh_gamma}
T_{\rm RH}= 
\begin{cases}
\l(\f{30}{\pi^2\,g_{\ast,{\rm RH}}}\r)^{1/4}  
\l(\f{n+4-6nl}{2n}\f{\Mpl^{4l-1}}{\sqrt{3}\gamma_\phi}\r)^{\f{1}{2(2l-1)}}\,& \text{for } 4+n-6nl>0,\label{eq:trh_gamma1}\\
\l(\f{30}{\pi^2\,g_{\ast,{\rm RH}}}\r)^{1/4}
\rho_{\rm end}^{\f{6nl-n-4}{8(n-2)}}
\l(\f{6nl-n-4}{2n}\f{\Mpl^{4l-1}}{\sqrt{3}\gamma_\phi}\r)^{-\f{3n}{4(n-2)}}\,& \text{for } 4+n-6nl<0.
\label{eq:trh_gamma2}
\end{cases}
\end{eqnarray}
\end{widetext}
Thus, by comparing Eqs.~\eqref{eq:Ninf_reheating} and \eqref{eq:trh_gamma}, one can establish a connection between the coupling parameters and the inflationary parameters, which will be comprehensively analyzed in Sec.~\ref{sec:result}. The reheating temperature, in principle, can be directly probed through the observations of the PGW background. However, the constraints on PGWs arises from the bounds on $\Delta N_{\rm eff}$ of the Universe, impose limits the measurement of the reheating temperature. These constraints will be explored in detail in the following section.
\begin{figure}[!ht]
    \centering
    \includegraphics[scale=0.4]{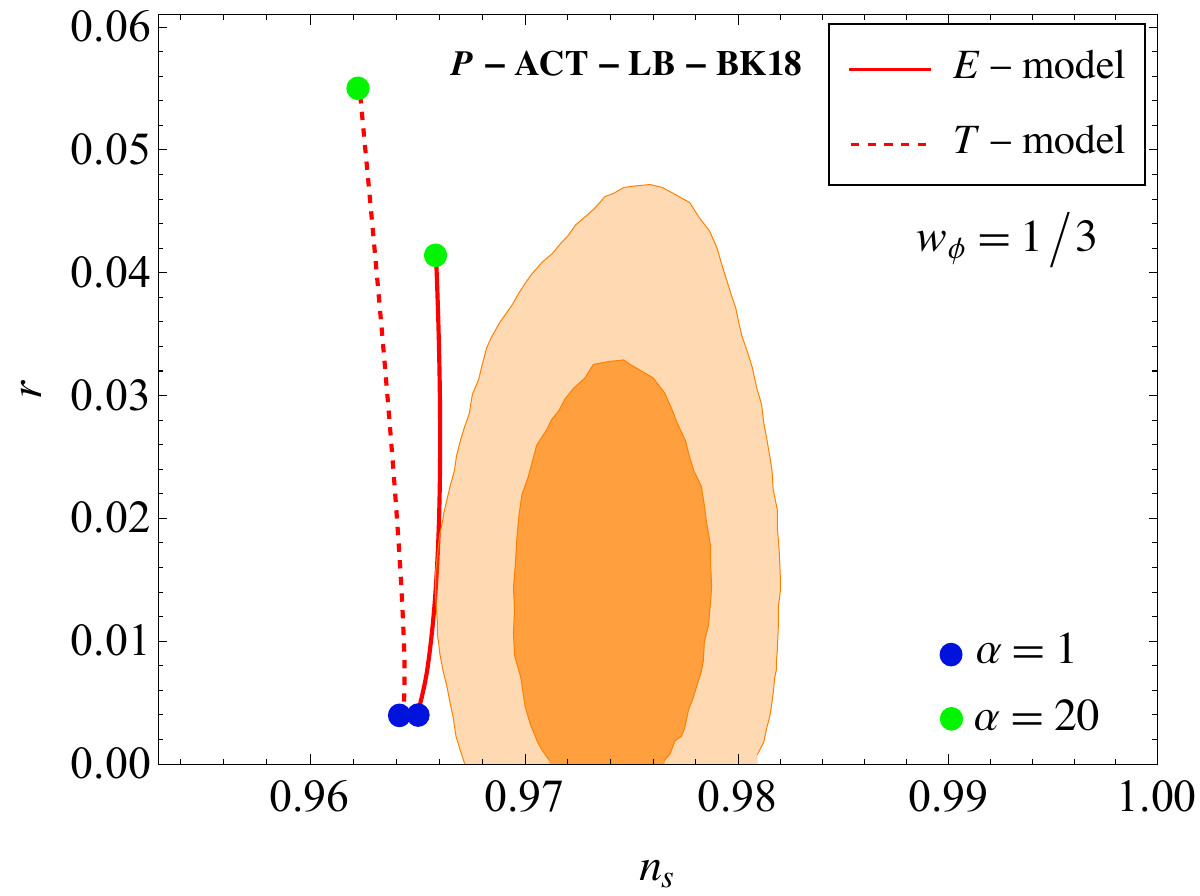}
    \caption[]{\justifying \it Predictions of both $\alpha$-attractor E- and T-model for $\w=1/3$ are shown projected onto the $(n_s, r)$ plane, overlaid with the latest combined constraints from {\rm P-ACT-LB-BK18}. The deep and light orange shaded regions indicate the $1\sigma$ (68\% C.L.) and $2\sigma$ (95\% C.L.) confidence intervals, respectively.}
    \label{fig:w13}
\end{figure}

\section{\texorpdfstring{$\Delta N_{\rm eff}$}{} Constraints from PGWs}
\label{sec:delta_neff}

In this section, we explore the constraints arising from spectral energy density of PGWs which are inherently generated by first-order tensor perturbations during the inflationary era, within the minimal setup \textit{i.e.}, in absence of any source terms. Thus, the signatures of the PGWs can not be ignored. As already mentioned, the primary observational constraint on PGWs comes from measurements of the tensor-to-scalar ratio $r$ at the pivot scale $k_{\ast}$, derived from CMB data . In $\alpha$-attractor models, the PGW spectrum is found to be nearly scale-invariant, as shown in detail in \cite{Caprini:2018mtu}. The amplitude and shape of the PGW spectral energy density are primarily determined by the energy scale of inflation.

To predict the present-day spectral energy density of modes that re-enter the horizon during the reheating era, one must consistently track the evolution of individual perturbation modes through the reheating phase following inflation. The inflation scale is directly related to $r$ and the amplitude $A_{\mathcal{R}}$ of the scalar perturbation.
Focusing on modes ($k<k_{\rm RH}$) that re-enter the horizon during radiation domination, the present-day dimensionless spectral energy density of PGWs is given by,
\begin{eqnarray}\label{eq:omega_gw_rad}
\Omega_{\rm GW}^{\rm (0)}(k)\, h^2
&= &\Omega^{\rm (0)}_{\rm R}\,h^2\,\frac{H_{\rm end}^2}{12\, \pi^2\, M_{\rm P}^2} \nonumber \\
&=& 3.5 \times 10^{-17}\, \left(\frac{H_{\rm end}}{10^{-5} M_{\rm P}}\right)^2,
\end{eqnarray}
where we have used the present-day radiation density $\Omega^{\rm (0)}_{\rm R}\,h^2=4.16 \times 10^{-5}$ including contributions from both photons and three neutrino species. If reheating phase precedes radiation domination, then modes in the intermediate range, $k_{\rm RH}<k<k_{\rm end}$ re-enter the horizon during reheating. In general, $k_{\rm RH}$ is determined by the EoS during reheating ($w_{\phi}$) and the reheating temperature ($T_{\rm RH}$), whereas the wavenumber $k_{\rm end}$ is primarily governed by $H_{\rm end}$. Thus, the GW spectral energy density accounting the effects of perturbation modes that re-enter the Hubble radius during the epoch of reheating is given by (see Ref.~\cite{Haque:2021dha} for detailed calculations),
\begin{align}
\Omega_{\rm GW}^{\rm (0)}(k)\,h^2 
&\simeq \Omega^{\rm (0)}_{R}\,h^2\,\frac{H_{\rm end}^2}{12\, \pi^2\, {M_{\rm P}}^2}\, \frac{\mu(w_{\phi})}{\pi}\,
\left(\frac{k}{k_{\rm RH}}\right)^{n_{w_{\phi}}}\nonumber\\
&\simeq  3.5 \times 10^{-17}\, \left(\frac{H_{\rm end}}{10^{-5}\, {M_{\rm P}}}\right)^2\,
\frac{\mu(w_{\phi})}{\pi}\,\left(\frac{k}{k_{\rm RH}}\right)^{n_{w_{\phi}}},\label{eq:omega_gw_reheating}
\end{align}
where the quantity $\mu(w_{\phi})$ and the index~$n_{w_{\phi}}$ are given by,
\begin{eqnarray}
\mu(w_{\phi})&\equiv&(1+3\, w_{\phi})^{4/(1+3\,w_{\phi})}\, \Gamma^2\left(\frac{5+ 3\,w_{\phi}}{2+6\,w_{\phi}}\right), \nonumber \\ 
n_{w_{\phi}}&\equiv&-\frac{2\,(1-3\,w_{\phi})}{1+3\,w_{\phi}}.
\end{eqnarray}
While $\mu(w_{\phi}) \simeq {\mathcal O}(1)$ for $0\leq w_{\phi} \leq 1$, the tilt
$n_{w_{\phi}}$ becomes positive or negative depending on whether $w_{\phi}>1/3$ or $w_{\phi}<1/3$, and 
vanishes at radiation domination ($w_{\phi}=1/3$).
Moreover, the wave number $k_{\rm end}$ can be approximately expressed in terms of 
the inflationary energy scale $H_{\rm end}$ and the reheating parameters $T_{\rm RH}$ as,
\begin{eqnarray}   
k_{\rm end} =& a_{\rm end }\,H_{\rm end} 
\simeq  H_{\rm end}\, \frac{T_0}{T_{\rm RH}}\,
\left(\frac{43}{11\, g_{*S,\mathrm{RH}}}\right)^{1/3}\, \nonumber \\ 
& \times \left({\frac{\pi^2\,g_{*\mathrm{RH}}\,T_{\rm RH}^4}{90\,{M_{\rm P}^2 \,H_{\rm end}^2}}}\right)^{1/[3\,(1+w_{\phi})]}.
\label{eq:k_end}
\end{eqnarray} 
On the other hand, the effective number of additional relativistic degrees of freedom (which is generally quantified by $\Delta N_{\rm eff}$) at the epoch of BBN, places tighter constraints on the reheating temperature.  
High-frequency GWs behave like additional relativistic degrees of freedom and contribute to total $\Delta N_{\rm eff}$. However, $\Delta N_{\rm eff}$ is tightly constrained by the recent ACT data, $\Delta N_{\rm eff} \leq 0.17$ with $95 \%$ C.L.\cite{ACT:2025fju,ACT:2025tim}. Following \cite{Chakraborty:2023ocr,Maity:2024odg}, the constraint equation on PGW in terms of $\Delta N_{\rm eff}$ is given by,
\begin{align}
\int_{k_{\rm RH}}^{k_{\rm end}}\frac{\d k}{k}\,\Omega^{\rm (0)}_{\rm GW}(k)\,h^2
\leq \frac{7}{8}\,\left(\frac{4}{11}\right)^{4/3}\,\Omega^{\rm (0)}_{\gamma}\,h^2\,\Delta N_\mathrm{eff},
\label{eq:deltaneff}
\end{align}
where $\Omega^{\rm (0)}_{\gamma}\,h^2\simeq 2.47\times10^{-5}$ represents the present-day energy density of photons. The above constraint is of particular interest whenever $w_{\phi}>1/3$, as spectral energy density of PGW has positive value  for modes with $k>k_{\rm RH}$.
Upon using the form of $\Omega^{\rm (0)}_{\rm GW}$ in Eq.~\eqref{eq:omega_gw_reheating}, the above condition reduces to,
\begin{eqnarray}
\Omega^{\rm (0)}_{\rm R}\,h^2\,\frac{H_{\rm end}^2}{12\, \pi^2\, M_{\rm P}^2}\, \frac{\mu(w_{\phi})\,(1+3\,w_{\phi})}{2\,\pi\,
(3\,w_{\phi} -1)}\,  \left(\frac{k_{\rm end}}{k_{\rm RH}}\right)^\frac{6\,w_{\phi} -2}{1+3\,w_{\phi}}  \nonumber \\ 
\leq 5.61\times 10^{-6}\,\Delta N_{\rm eff} \, .\label{Eq:BBNapprox}
\end{eqnarray}
The ratio between $k_{\rm end}$ and $k_{\rm RH}$ in the above equation can be further expressed as,
\begin{equation}
\frac{k_{\rm end}}{k_{\rm RH}}
=\left(\frac{90\,H_{\rm end}^2\,{M_{\rm p}}^2}{\pi^2\,g_{* \mathrm{RH}}}\right)^\frac{(1+3\,w_{\phi})}{[6\,(1+w_{\phi})]}\,
T_{\rm RH}^{-2\,(1+3\,w_{\phi})/[3\,(1+w_{\phi})]} \,. \label{eq:ke}
\end{equation}
Combining Eqs.~\eqref{eq:deltaneff}, \eqref{Eq:BBNapprox} and \eqref{eq:ke}, one can arrive at the following bound on the reheating temperature $T_{\rm RH}$,
\begin{eqnarray}
T_{\rm RH} \geq & \left[\frac{\Omega^{\rm (0)}_{\rm R}\, h^2}{5.61\times 10^{-6}\,\Delta N_{\rm eff}}\,
\frac{H_{\rm end}^2}{12\, \pi^2\, {M_{\rm P}}^2}\, \frac{\mu(w_{\phi})\,(1+3\,w_{\phi})}{2\,\pi\,
(3\,w_{\phi} -1)}\,\right]^{\frac{3\,(1+w_{\phi})}{4\,(3\,w_{\phi} -1)}}\, \nonumber \\
 & \times \left(\frac{90\,H_{\rm end}^2\,{M_{\rm p}}^2}{\pi^2\,g_{*\mathrm{RH}}}\right)^{\frac{1}{4}} \equiv T_{\rm RH}^{\rm GW}.
\label{eq:BBNrestriction}
\end{eqnarray}
Taking $T_{\rm RH}^{\rm GW} \sim T_{\rm BBN} \approx 4\, {\rm MeV}$, we observe that the BBN constraint on PGWs becomes relevant only if $w_{\phi} \geq 0.6$. This threshold defines a new lower bound on the reheating temperature due to PGWs, denoted as $T_{\rm RH}^{\rm GW}$.
The above condition implies that for particular values of ($w_{\phi}$, $T_{\rm RH}$), the upper bound on $\Delta N_{\rm eff}$ constrains the energy scale of inflation or equivalently imposes a constraint on $r$. This further puts constraint on the coupling parameters. In the following, we will discuss how the inflaton–SM coupling modifies the effective inflationary potential through one-loop corrections. To ensure that inflation remains unaffected by these radiative effects, we will derive an upper bound on the coupling parameter using the one-loop radiative correction formalism \textit{in inflation potential}~\cite{Coleman:1973jx}.

\section{One-loop correction in inflation potential}
\label{sec:loop_correction}
\begin{figure*}
    \centering
    \includegraphics[scale=0.35]{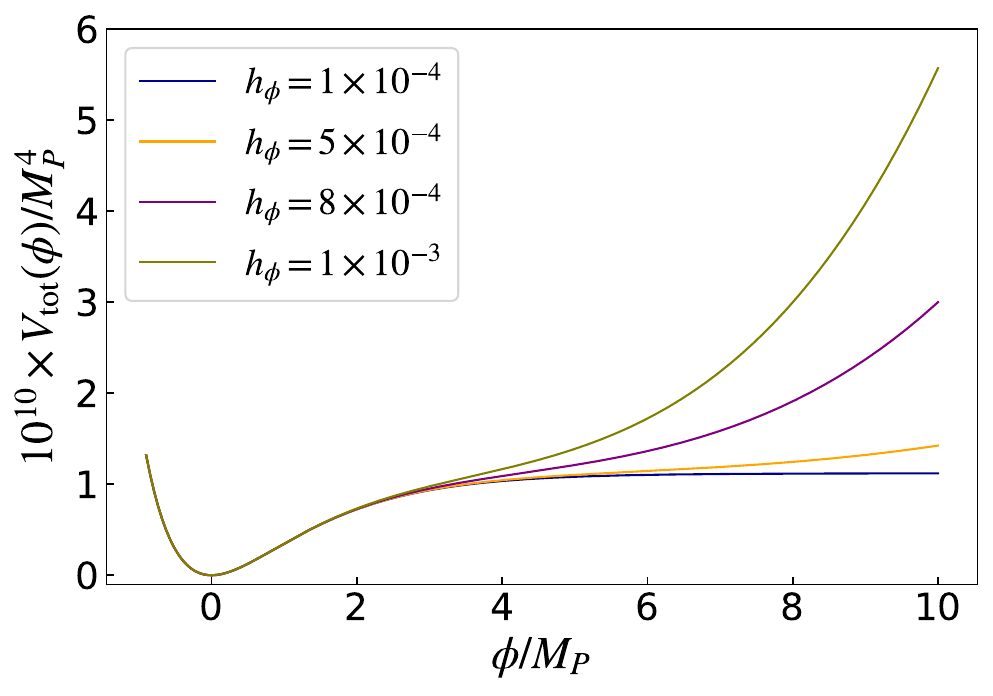}
    \hfill
    \includegraphics[scale=0.35]{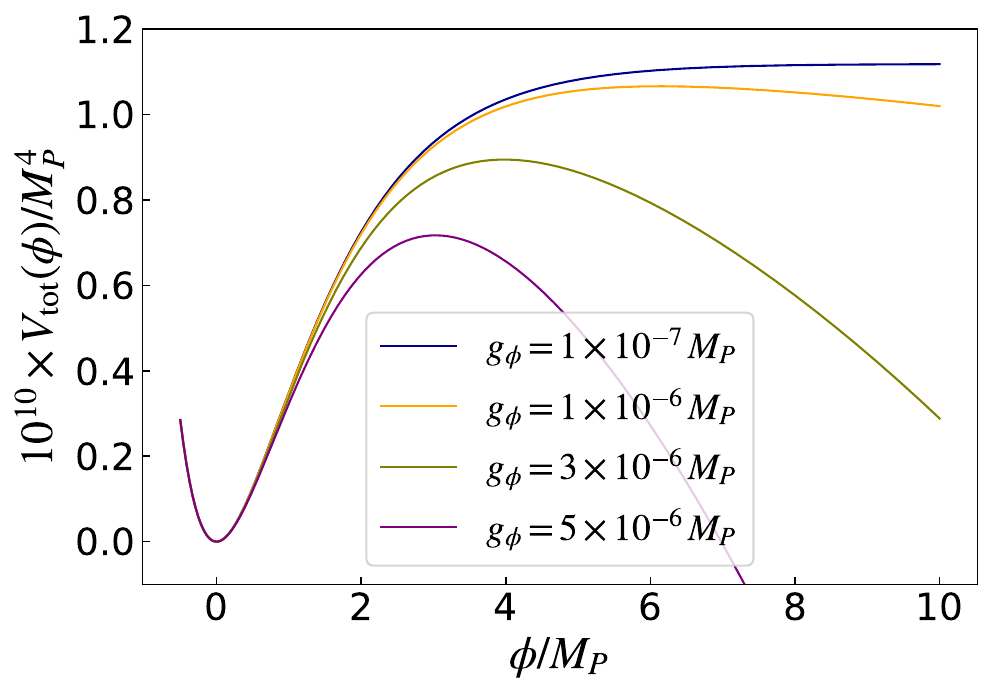}
    \hfill
    \includegraphics[scale=0.35]{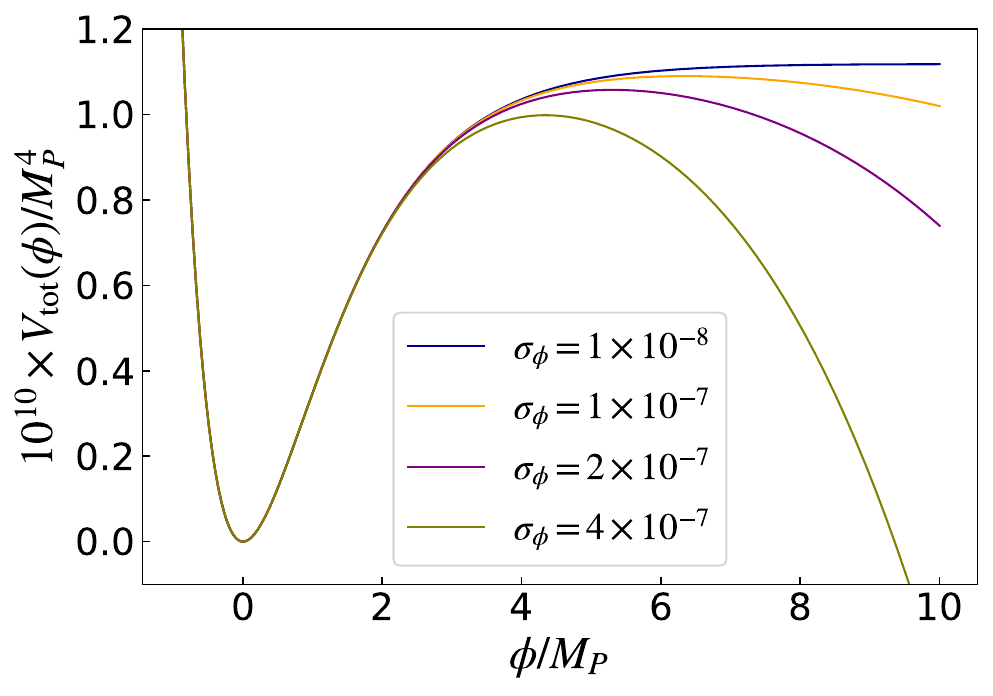}
    \hfill
    \includegraphics[scale=0.35]{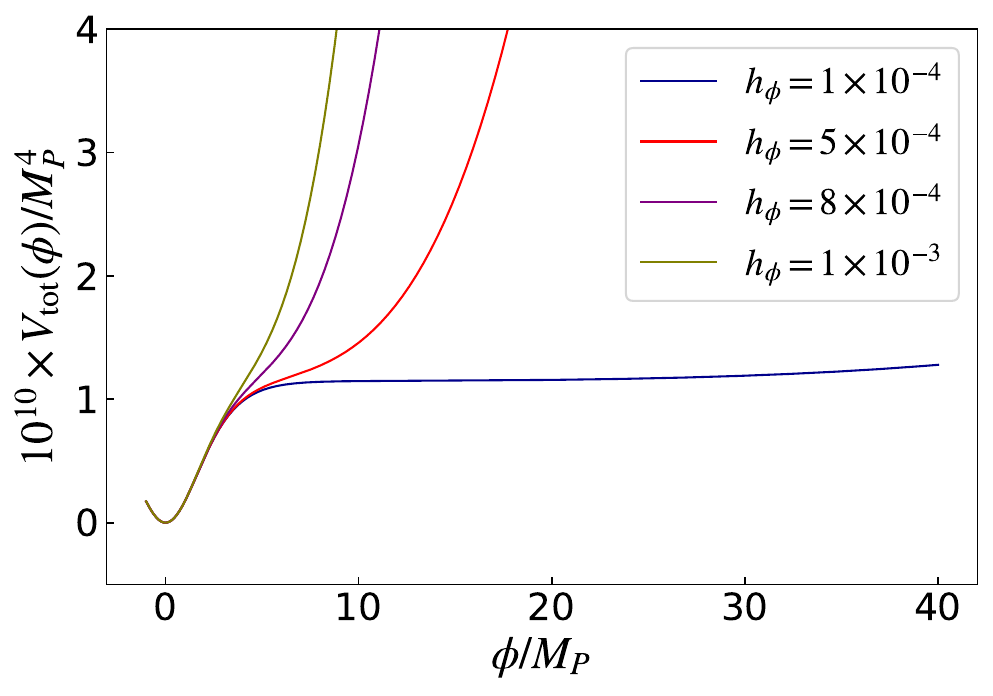}
    \hfill
    \includegraphics[scale=0.35]{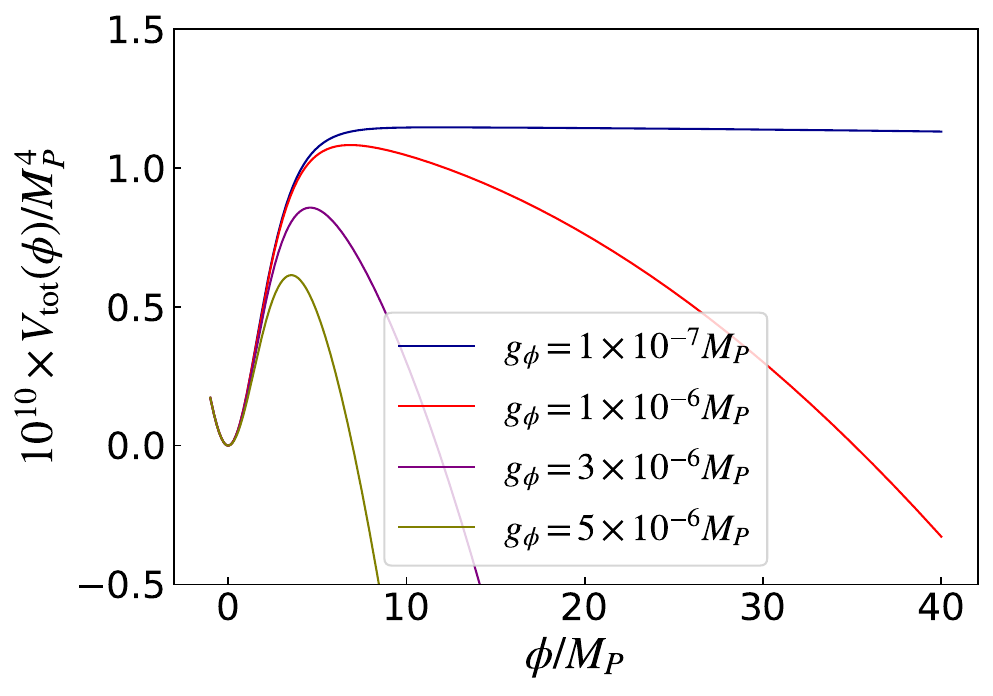}
    \hfill
    \includegraphics[scale=0.35]{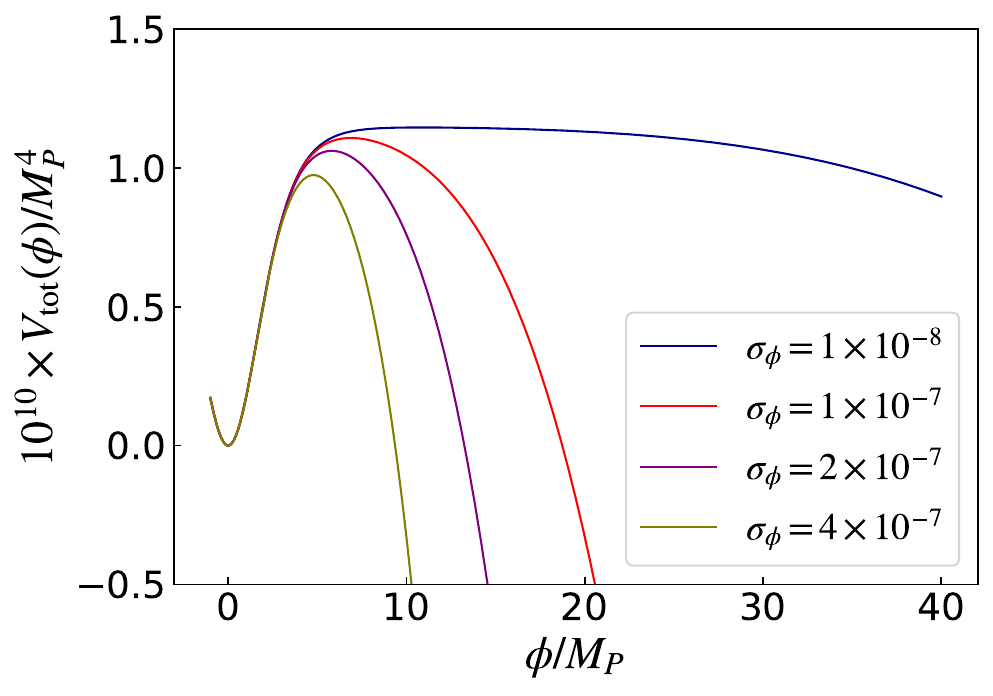}
    \caption[]{\justifying \textit{Coleman--Weinberg one-loop corrected inflationary potentials for the E-model (upper panel) and T-model (lower panel), assuming $n=1$ and $\alpha=1$. The left, middle, and right panels show the dependence of the corrected potential on the coupling parameter for the decay/annihilation channels $\phi \to \bar{f}f$, $\phi \to bb$, and $\phi\phi \to bb$, respectively.}}
    \label{fig: CW}
\end{figure*}
In the aforementioned sections, we have shown how the minimum reheating temperature is modified when accounting for the overproduction of PGWs in light of the $\Delta N_{\rm eff}$ bound. In this section, we examine how inflaton–SM couplings can significantly alter the effective inflationary potential through one-loop radiative corrections~\cite{Coleman:1973jx,Armendariz-Picon:2020tkc, Bastero-Gil:2016qru,Gorgulho:2025wxz}. These corrections impose an upper bound on the coupling parameters~\cite{Kallosh:2013tua,Kallosh:2013yoa,Wolf:2025ecy,Gialamas:2025kef}: beyond this bound, the radiative effects become large enough that our analysis is no longer reliable. To quantify this constraint, we employ the Coleman–Weinberg (CW) one-loop radiative correction formalism. Within this framework, the one-loop corrected inflationary potential is given by~\cite{Coleman:1973jx,Drees:2021wgd},
\begin{eqnarray}
    \label{eq:CW_potential}
    V_{\rm CW}(\phi) = \sum_i \frac{n_i}{64\pi^2}(-1)^{2s_i}m_i^4(\phi)\left[\ln\left(\frac{m_i^2(\phi)}{\mu^2}\right) - \frac{3}{2}\right],
\end{eqnarray}
with the summation running over the radiation fields ($f,b$) which is coupled to inflaton. Here, $s_i$ represents the spin of the fields and $n_i$ is the total internal degrees of freedom. $\mu$ is a renormalization scale which is considered to be $\phi_k$. $m_i$ represents the $\phi$-dependent mass term for the radiation fields, which can be expressed for different interactions as
\begin{eqnarray}
    m_i^2(\phi) \equiv 
    \begin{cases}
        y_{\phi}^2 \phi^2       \qquad\,{\rm for}\quad \phi \bar{f}f \\
        2g_{\phi} \phi          \qquad\,{\rm for}\quad \phi bb \\
        2\sigma_{\phi} \phi^2   \,\,\,\quad\,{\rm for}\quad \phi \phi bb.
    \end{cases}
\end{eqnarray}
In Fig.~\ref{fig: CW}, we show that the one-loop correction can significantly affect the inflationary potential depending on the coupling strength. The figure illustrates how the effective E-model (upper panel) and T-model (lower panel) potentials, with $\alpha=1$ and $n=1$, are modified by different non-gravitational interactions (\textit{cf.} Eq.~\eqref{eq:Lint}) through the CW one-loop correction. To ensure the stability of the inflationary potential, the CW-corrected potential must satisfy the following inequality at $\phi=\phi_k$:
\begin{eqnarray}
    V_{\rm CW} (\phi_k) < \left|V_{\rm tree} (\phi_k)\right|,
    \label{eq:CW_inequality}
\end{eqnarray}
where $\left|V_{\rm tree}(\phi)\right|$ is the effective tree potential as defined in Eq.~\eqref{eq:attractorpotential}. This inequality yields an upper limit on the coupling parameters, given by
\begin{eqnarray}
    y_{\phi} &<& \exp\left[\frac{3}{4}+ \frac{1}{4}\mathcal{W}_{-1}\left(-\frac{64\pi^2\left|V_{\rm tree}(\phi_k)\right|}{e^3\phi_k^4}\right)\right]\,, \nonumber\\
    \frac{g_{\phi}}{M_{\rm P}} &<& \frac{\phi_k}{2M_{\rm P}}\exp\left[\frac{3}{2}+ \frac{1}{2}\mathcal{W}_{-1}\left(\frac{128\pi^2\left|V_{\rm tree}(\phi_k)\right|}{e^3\phi_k^4}\right)\right]\,, \nonumber\\
    \sigma_{\phi} &<& \frac{1}{2}\exp\left[\frac{3}{2}+ \frac{1}{2}\mathcal{W}_{-1}\left(\frac{128\pi^2\left|V_{\rm tree}(\phi_k)\right|}{e^3\phi_k^4}\right)\right], \label{eq:coupling_CW}
\end{eqnarray}
where $\mathcal{W}_{-1}$ is the Lambert function of branch $-1$.

The CW correction plays a crucial role in determining the validity of our results. In particular, it can significantly modify the inflationary potential depending on the coupling strength, thereby altering the parameter space inferred from Figs.~\ref{fig:emodel_refinement_w0}--\ref{fig:emodel_refinement}. The shaded regions in these figures indicate the parameter space where the CW correction becomes important, and hence our analysis can no longer be trusted.

In the following section, we analyze the impact of self-resonance during the post-inflationary dynamics, which can play an important role. This phenomenon indicates that homogeneous oscillations of the inflaton can become unstable to small spatial perturbations, leading to the parametric amplification of its own fluctuations. We show that this effect further constrains the viable parameter space of the model of interest.

\section{Impact of self-resonance during reheating phase}
\label{sec:self_resonance}

\begin{figure}
    \centering
    \includegraphics[scale=0.5]{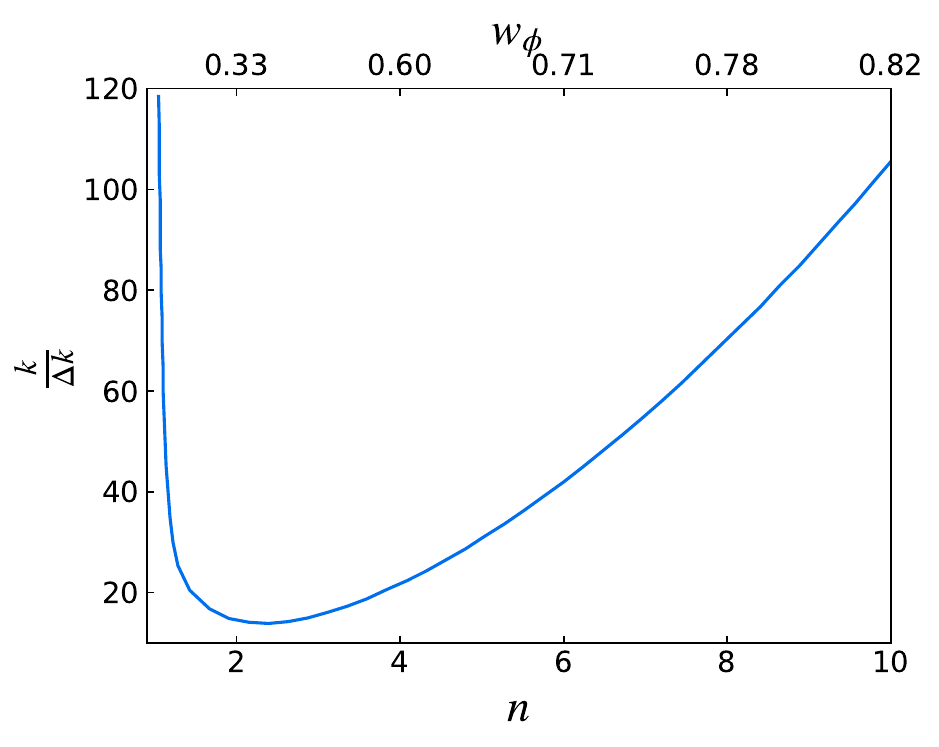}
    \caption[]{\justifying \it Variation of inverse fractional width with respect to $n$. The top axis presents $\w$ which is related to $n$ according to Eq.~\eqref{eq:EoS}. The figure is reproduced from Ref.~\cite{Lozanov:2017hjm}.}
    \label{fig:koverdeltak}
\end{figure}

\begin{figure*}
    \centering
    \includegraphics[scale=0.45]{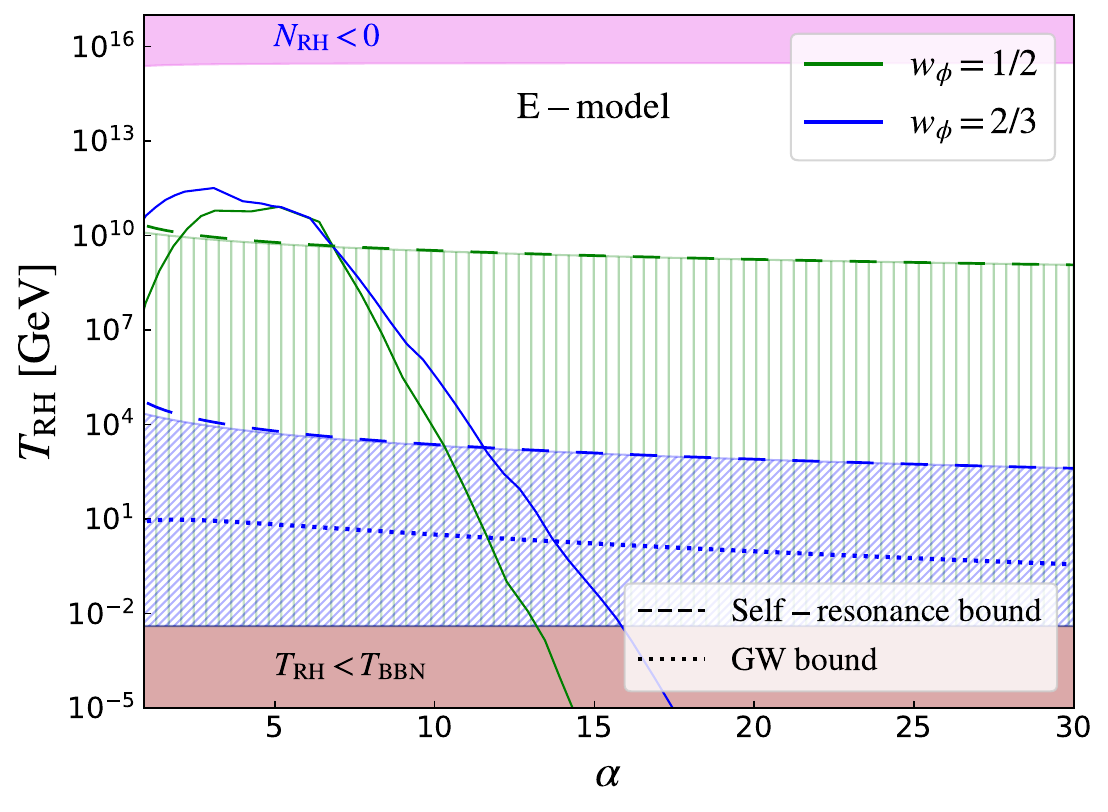}
    \includegraphics[scale=0.45]{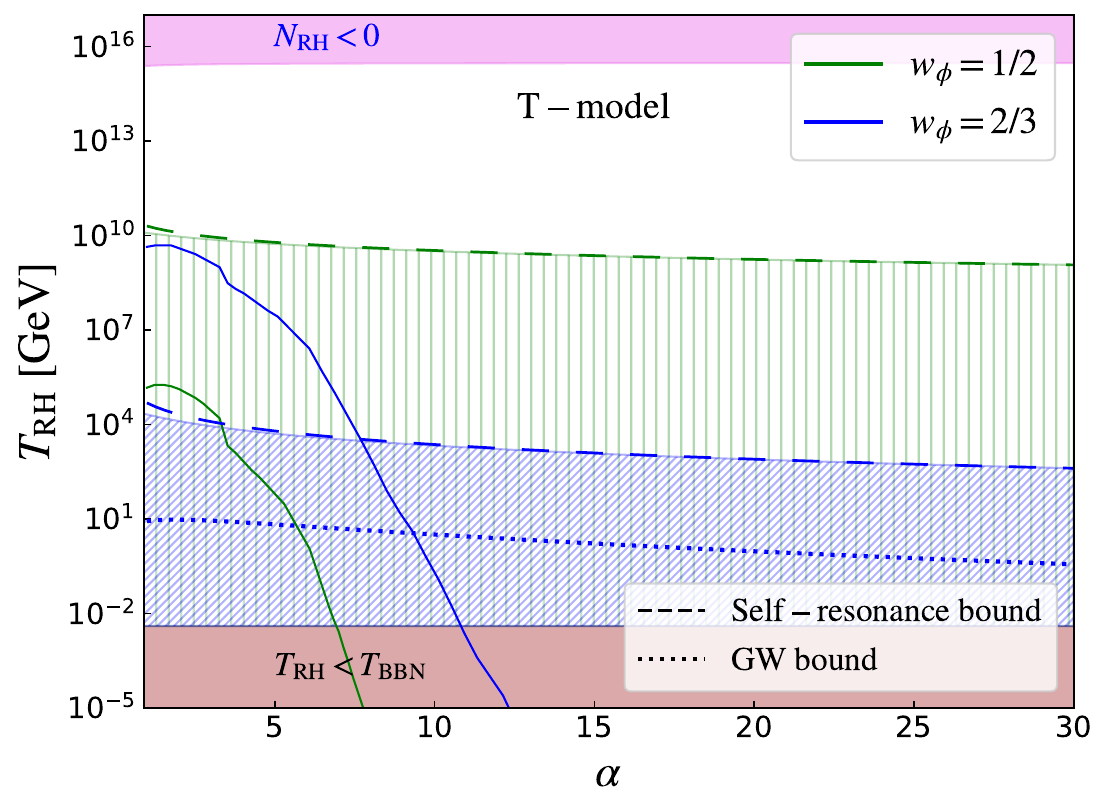}
    \caption[]{\justifying\it The $2\sigma$-consistent region of the reheating temperature in the $T_{\rm RH}$-$\alpha$ plane, considering recent P-ACT-LB-BK 18 data is shown by solid lines, with the enclosed area representing the allowed parameter space. On top of this, we impose additional consistency conditions from self-resonance and PGW overproduction bounds for both E- (left) and T-models (right). The magenta and brown bands denote excluded regions corresponding to $N_{\rm RH}<0$ and $T_{\rm RH}<T_{\rm BBN}$, respectively. The constraints from self-resonance and GW overproduction are indicated by dashed and dotted lines, respectively, with colors (green for $w=1/2$ and blue for $w=2/3$) representing different EoS. }
    \label{fig:trh_self_resonance}
\end{figure*}
In this section, we analyze the impact of self-resonance on post-inflationary dynamics. After inflation ends, the homogeneous inflaton field $\phi(t)$ oscillates around the minimum of its potential. However, small spatial perturbations can render these oscillations unstable, triggering self-resonance even in the absence of couplings to SM fields~\cite{Amin:2010dc,Amin:2010xe,Amin:2011hj}. This leads to the parametric amplification of inflaton fluctuations and enables efficient, non-perturbative energy transfer driven solely by the inflaton’s nonlinear dynamics. Following Ref.~\cite{Lozanov:2017hjm}, we estimates for the duration required for the EoS to approach $\w \to 1/3$ after inflation, denoted as $N_{\rm RH}^{\rm (sr)}$. Therefore, for any choice of coupling used in our perturbative reheating analysis, if the reheating duration satisfies $N_{\rm RH} > N_{\rm RH}^{\rm (sr)}$, our results cannot be considered reliable, as self-resonance alone is sufficient to reheat the Universe.

For an inflaton potential of the form $V(\phi)\propto \phi^{2n}$, the number of reheating $e$-folds in the self-resonant reheating scenario is given by by~\cite{Lozanov:2016hid,Lozanov:2017hjm}
\begin{eqnarray}
    \label{eq:Nrh_selfresonance}
    N_{\rm RH}^{\rm (sr)} \simeq 
    \begin{cases}
        \frac{n+1}{3}\ln\left[\frac{\sqrt{6\alpha}}{\delta d}\frac{k}{\Delta k}\frac{|4-2n|}{n+1}\right]&\quad {\rm for }\,\, n>1,n\neq2\\
        \ln\left[\frac{\sqrt{6\alpha}}{\delta d}\right]&\quad {\rm for }\,\, n=2,
    \end{cases}
\end{eqnarray}
where $\frac{k}{\Delta k}$ is the inverse fractional width of the resonance band and $d$ is its dimensionless `strength' which is $d\approx \frac{\Delta k}{k}$~\footnote{For $n=1$, self-resonance alone is not sufficient to reheat the Universe to a radiation-dominated state consistent with successful BBN. This requires additional couplings of the inflaton to SM fields to enable the eventual decay of inflaton into SM particles ~\cite{Lozanov:2017hjm}.}. The variation of the inverse fractional width with respect to EoS, in presented in Fig.~\ref{fig:koverdeltak}. The parameter $\delta\sim0.126$ is independent of $n$.

A complete treatment of self-resonance requires a comprehensive non-perturbative analysis of the inflaton’s nonlinear dynamics and fragmentation, including mode amplification and backreaction effects, which lies beyond the scope of the present work and will be addressed in future studies. In this paper, we therefore focus on the perturbative reheating regime, valid as long as $N_{\rm RH}<N_{\rm RH}^{\rm (sr)}$. Thus, this inequality provides a lower limit on the reheating temperature according to Eq.~\eqref{eq:trh_w_nrh}, as depicted in Fig.~\ref{fig:trh_self_resonance}. Table~\ref{tab:model_constraint_self_resonance} summarizes the updated limits on the inflationary parameters ($\alpha$, $N_k$) and the corresponding bounds on the reheating temperature as well as on the different coupling strength  after incorporating the self-resonance constraint. We find that these bounds are crucial for our analysis, as they delineate the region of parameter space where the perturbative treatment remains valid (see, for instance, Fig.~\ref{fig:trh_self_resonance} and \ref{fig:emodel_refinement}).

\section{Results}
\label{sec:result}
\begin{figure*}
    \centering
    \includegraphics[scale=0.45]{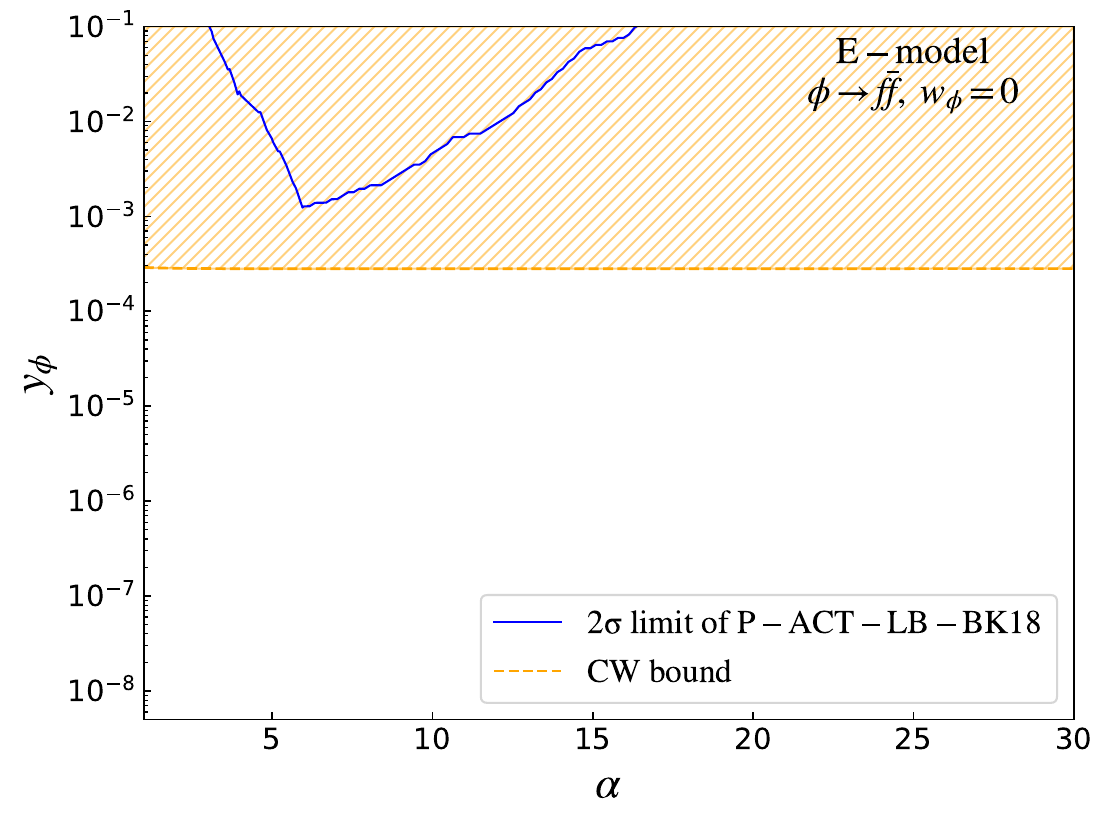}
    \includegraphics[scale=0.45]{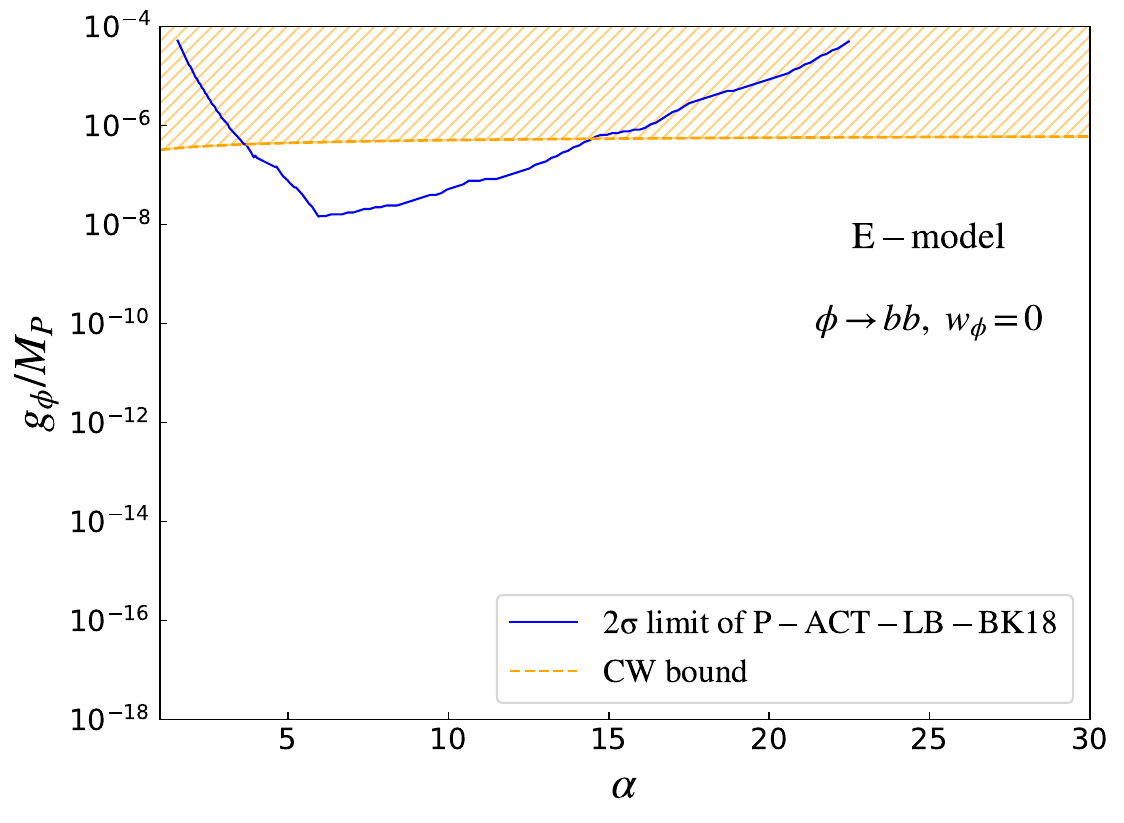}
    \caption[]{\justifying\it Constraints on different inflaton-SM coupling scenarios for the \textbf{E-model} of inflation, derived using the $2\sigma$ bounds from recent {\rm P-ACT-LB-BK18} data, shown as a function of $\alpha$ for $w=0$. The upper limits on the couplings arise from the CW one-loop correction. The left panel corresponds to the $\phi \to f\bar{f}$ decay channel, while the right panel corresponds to the $\phi \to bb$ decay channel. The case $w=0$ is not viable for the $\phi\phi \to bb$ scenario, as reheating cannot be achieved.}
    \label{fig:emodel_refinement_w0}
\end{figure*}
This section presents the observational constraints on the $\alpha$-attractor inflationary models and their associated reheating dynamics. We focus on how the inflationary observables ($n_s$, $r$) are related to the number of e-foldings $N_k$, which in turn depends on the reheating temperature $T_{\text{RH}}$ and $w_\phi$. These observables are expressed in terms of the inflationary model parameters $\alpha$ and $n$, establishing a direct link between the inflationary dynamics and post-inflationary reheating. In particular, since the number of e-foldings is sensitive to the thermal history following inflation, we use Eqs.~\eqref{eq:Nk} and \eqref{eq:Ninf_reheating} simultaneously to evaluate $n_s$ in terms of $\alpha$, $n$ and $T_{\rm RH}$, thereby enabling constraints on these parameters using current data.

Furthermore, the reheating temperature is connected to the inflaton–SM coupling via Eq.~\eqref{eq:trh_gamma} and, allowing for a joint constraint on both the inflationary potential and the coupling strength. Our analysis incorporates the most recent observational data, including Planck18 with lensing, ACT DR6, BAO from DESI, and BICEP/$Keck$ 2018 (collectively denoted as P-ACT-LB-BK18)~\footnote{In particular, $1\sigma$ and $2\sigma$ bounds are taken from figure 10 of~\cite{ACT:2025tim}.  }. These datasets provide tight constraints on $n_s$ and $r$, which in turn translate into bounds on the reheating temperature and the microphysics of reheating \textit{i.e.}, the coupling between the inflaton and SM particles.

In addition to CMB constraints, we incorporate bounds from the overproduction of PGWs generated during inflation, which are constrained by the effective number of relativistic species, $\Delta N_{\rm eff}$, during BBN. Since PGW amplitudes are sensitive to the post-inflationary expansion history, particularly in stiff EoS scenarios, reheating can enhance GWs at small scales. Imposing $\Delta N_{\rm eff}$ bounds thus sets a lower limit on the reheating temperature and can significantly restrict the parameter space. This bound can exclude the entire $1\sigma$ region allowed by P-ACT-LB-BK18, leaving only a narrow viable band at $2\sigma$, which is otherwise consistent as shown in Refs.~\cite{Drees:2025ngb,Zharov:2025evb,Liu:2025qca}. Throughout this work, we adopt the conservative lower bound $T_{\rm RH} > T_{\rm BBN} \approx 4\,\mathrm{MeV}$ \cite{Kawasaki:1999na,Kawasaki:2000en,Hasegawa:2019jsa}, ensuring consistency with standard cosmological evolution.

Apart from the constraints discussed above, large inflaton--SM couplings can significantly modify the effective inflationary potential through one-loop radiative corrections. In particular, sizable couplings can induce corrections to the tree-level potential, thereby altering the inflationary dynamics and the predicted values of $n_s$ and $r$. To ensure the consistency of the inflationary framework, we incorporate the CW one-loop formalism to evaluate these effects. Requiring that the radiative corrections remain subdominant compared to the tree-level potential allows us to place an upper bound on the inflaton--SM couplings.

On the other hand, even in the absence of direct couplings to SM fields, small spatial fluctuations in the inflaton condensate during the post-inflationary oscillatory phase can render the homogeneous mode unstable. This instability can trigger inflaton self-resonance, leading to efficient energy transfer through nonlinear dynamics alone. In such a scenario, reheating proceeds via non-perturbative processes, necessitating a dedicated analysis beyond the perturbative decay framework. Since self-resonance can occur even for negligible inflaton--SM couplings, it cannot be ignored. Therefore, in the present work, we restrict ourselves to the perturbative reheating regime, characterized by the condition that self-resonance remains subdominant, $N_{\rm RH} < N_{\rm RH}^{\rm (sr)}$ (see Eq.~\eqref{eq:Nrh_selfresonance}). This requirement translates into a lower bound on the reheating temperature and, correspondingly, on the inflaton--SM couplings, ensuring that reheating proceeds predominantly through perturbative inflaton decay.


\begin{figure*}
    \centering
    \includegraphics[scale=0.3]{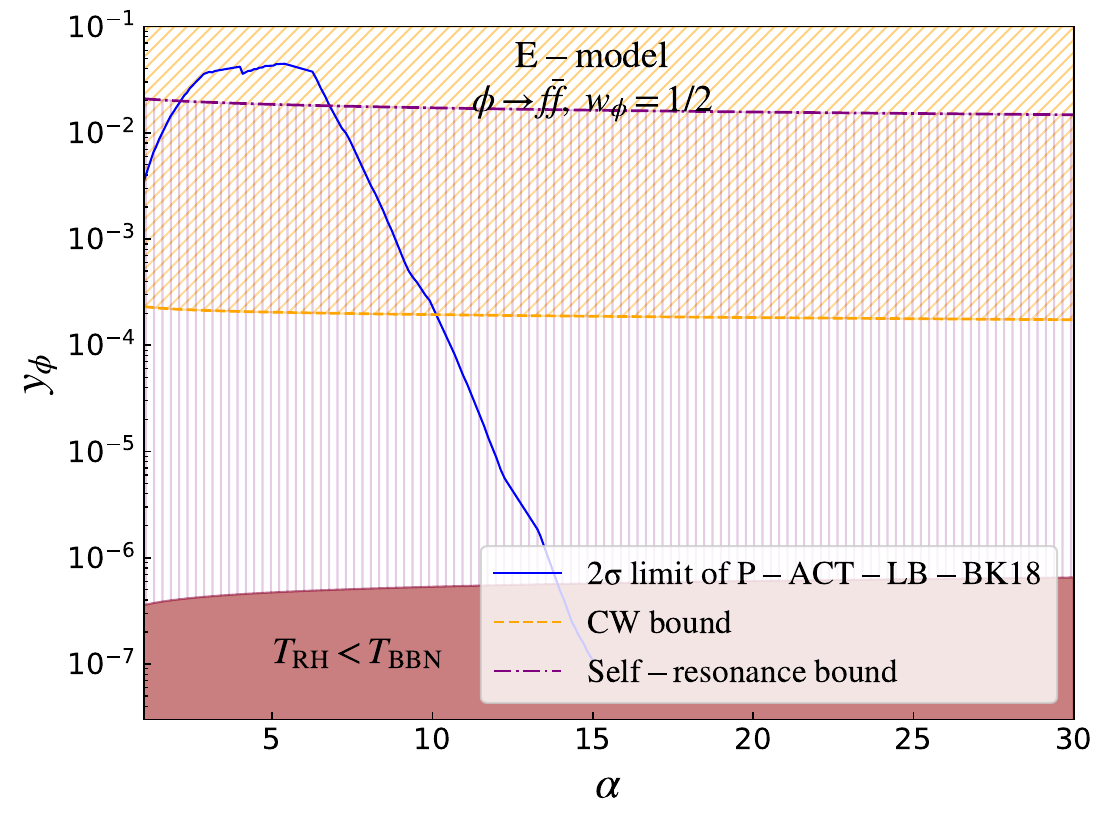}
    \includegraphics[scale=0.3]{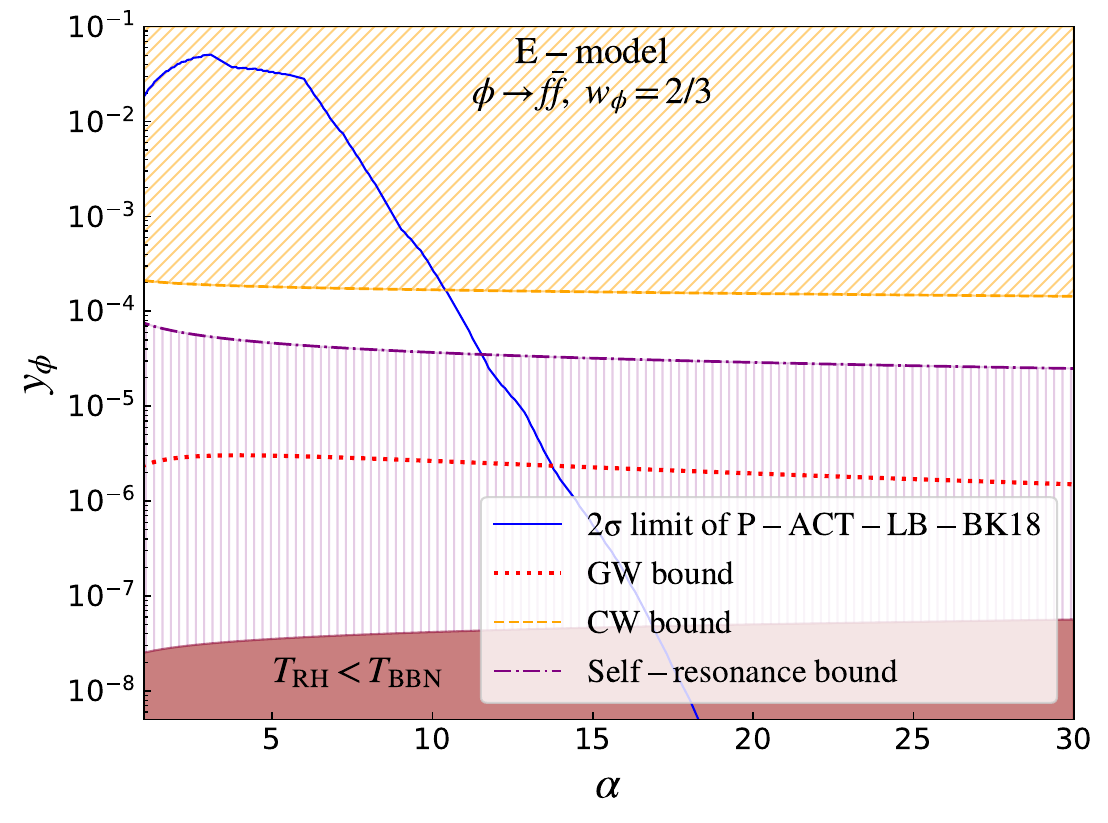}
    \includegraphics[scale=0.3]{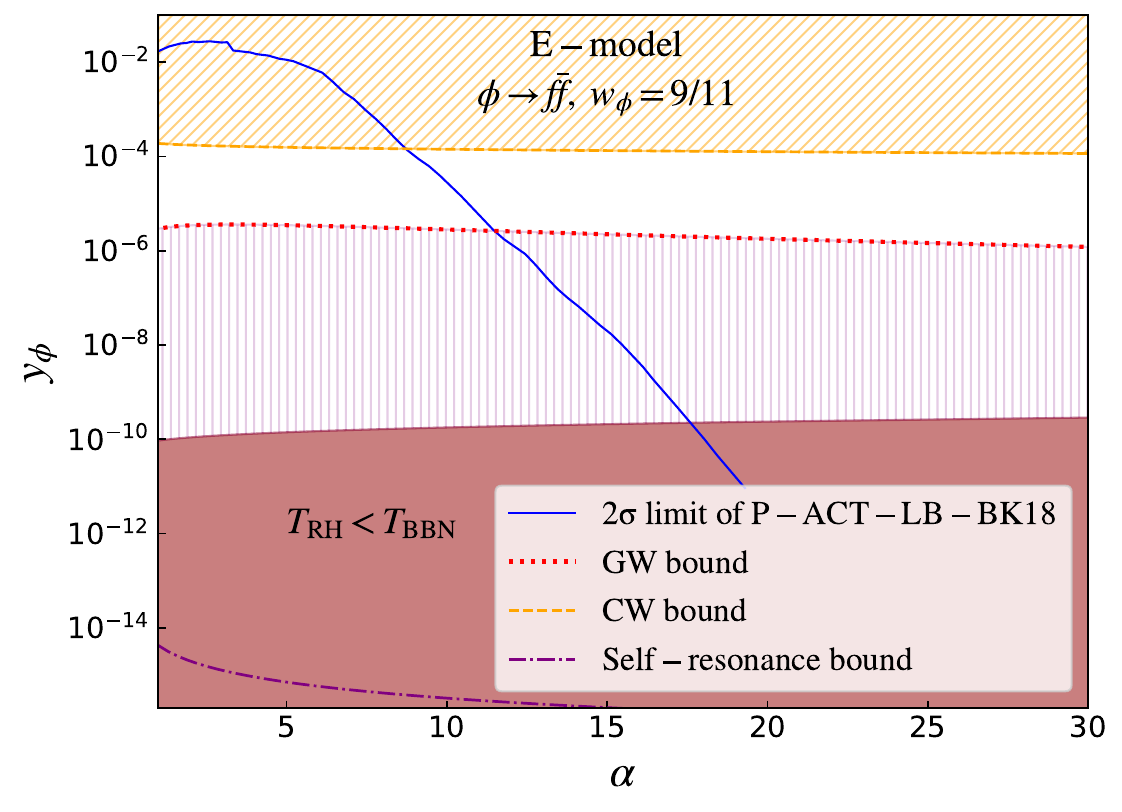}
    \includegraphics[scale=0.3]{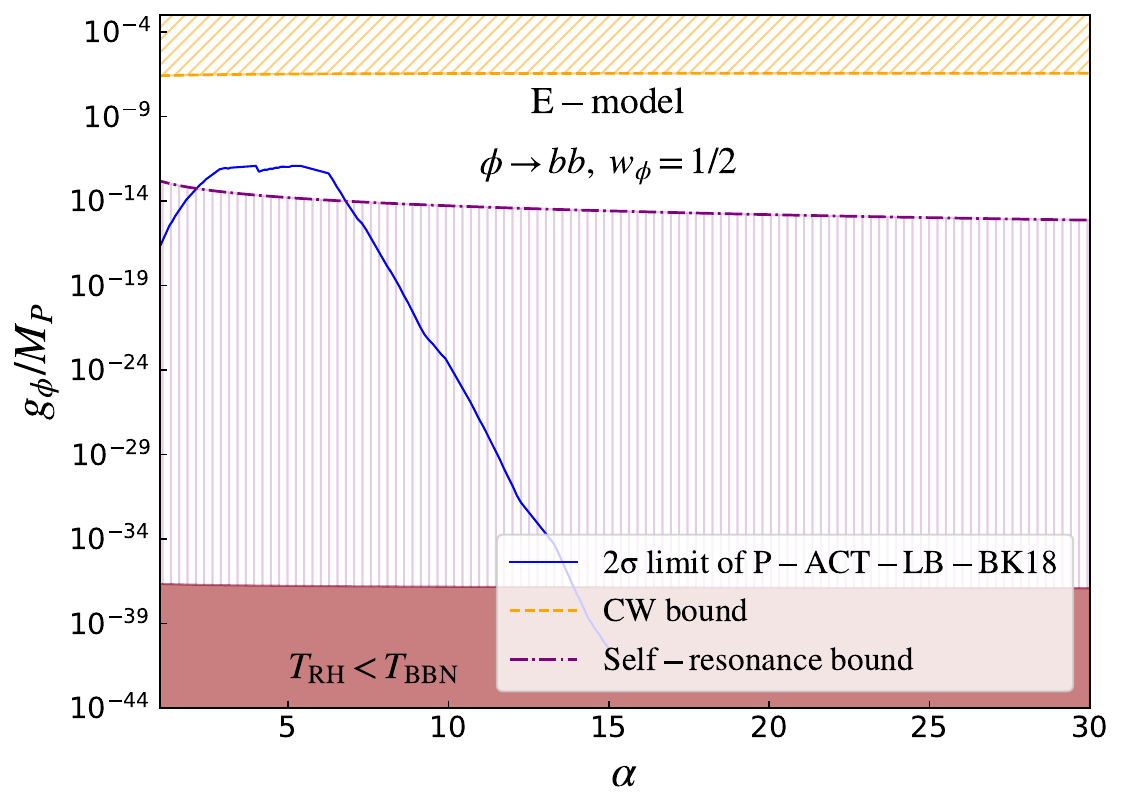}
    \includegraphics[scale=0.3]{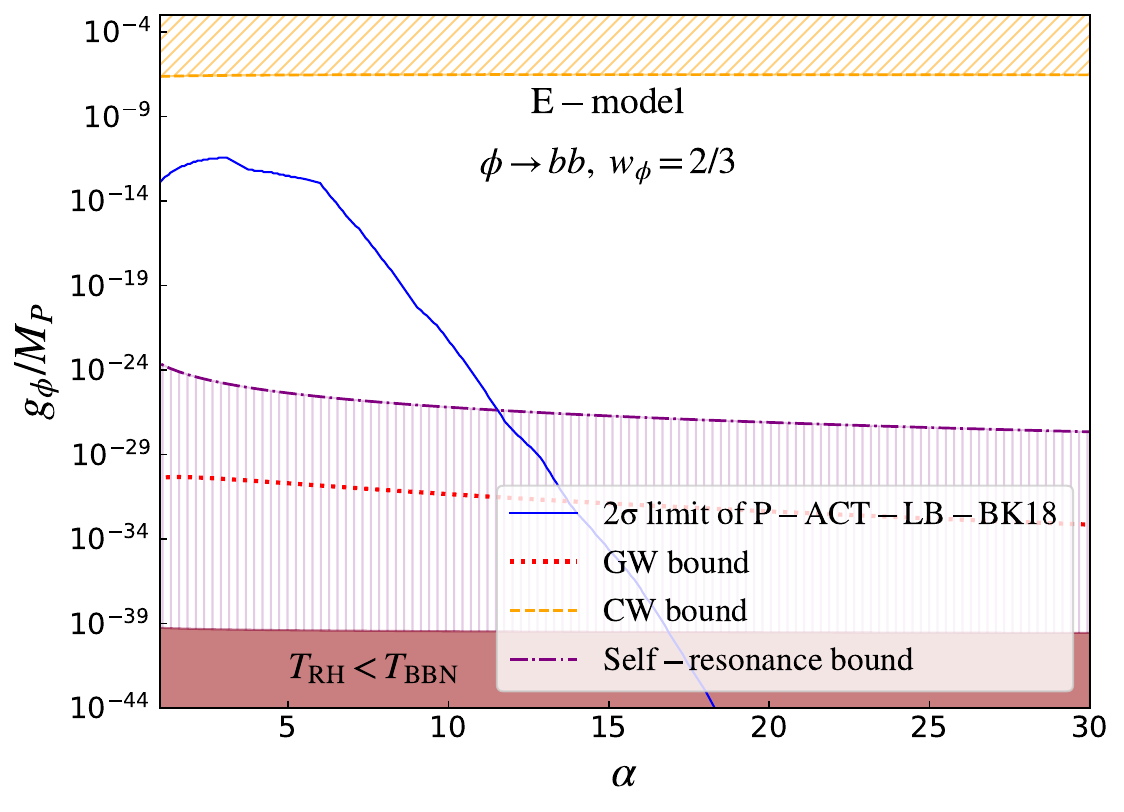}
    \includegraphics[scale=0.3]{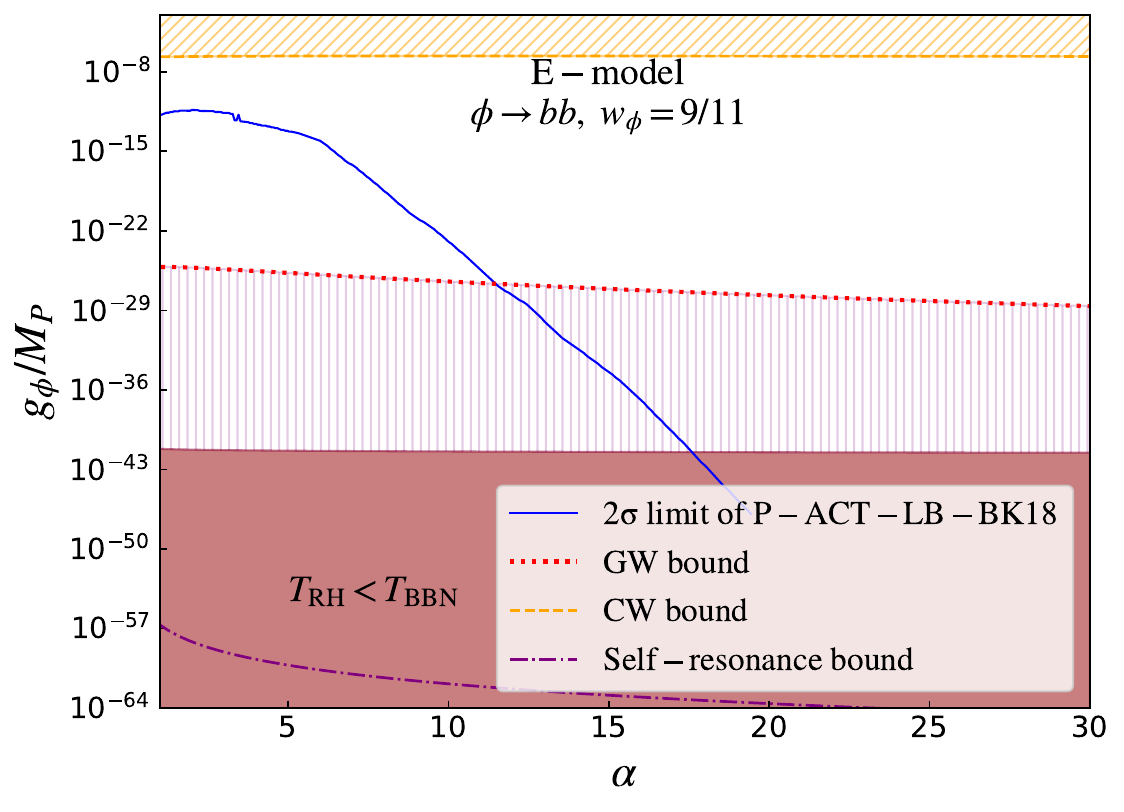}
    \includegraphics[scale=0.3]{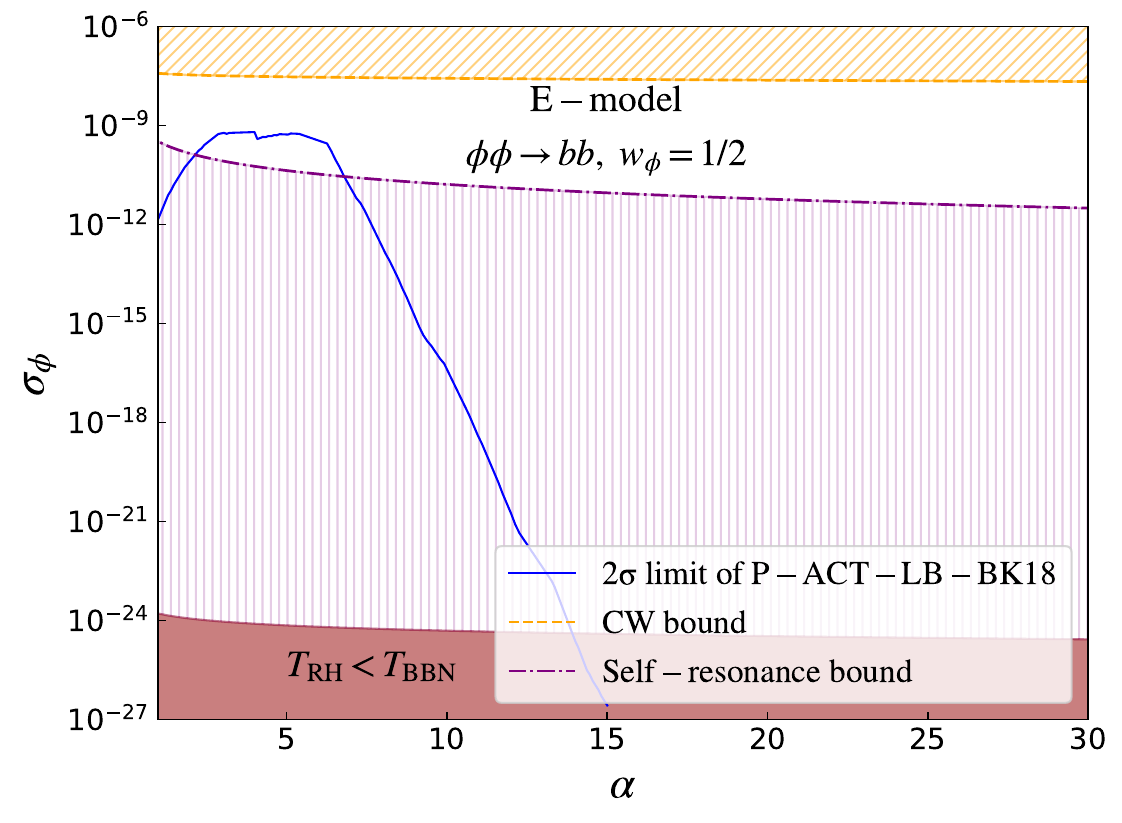}
    \includegraphics[scale=0.3]{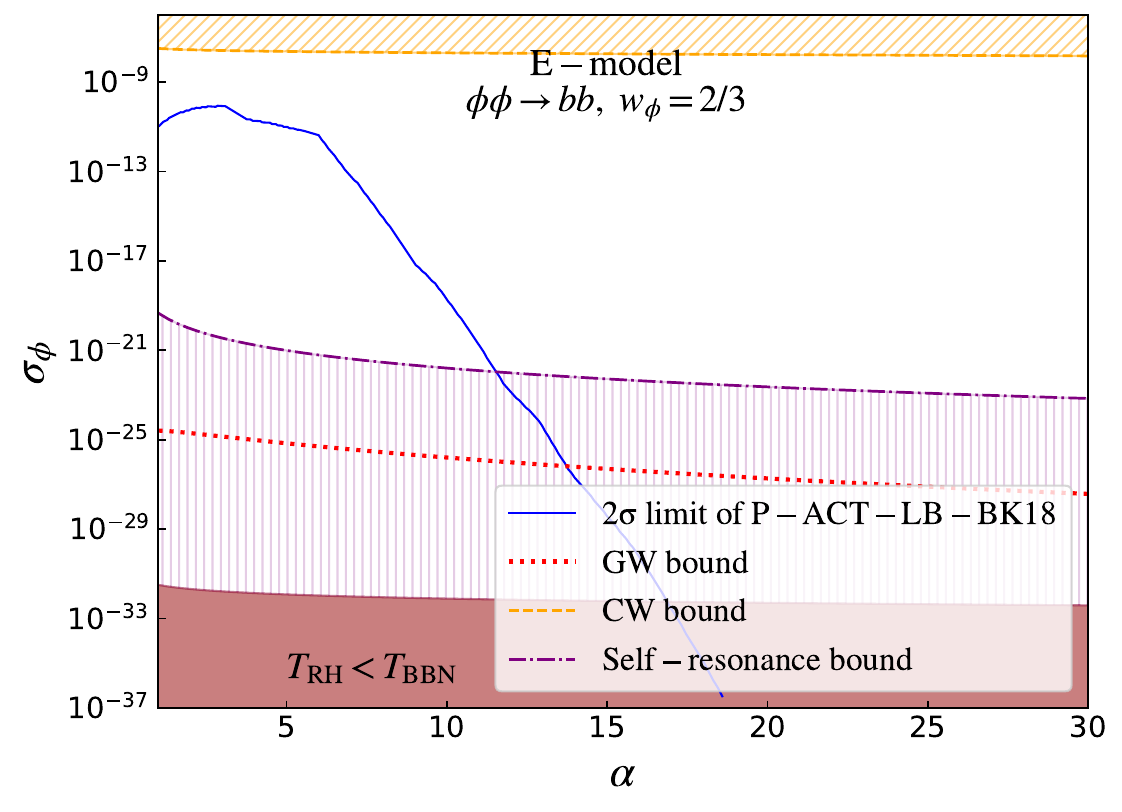}
    \includegraphics[scale=0.3]{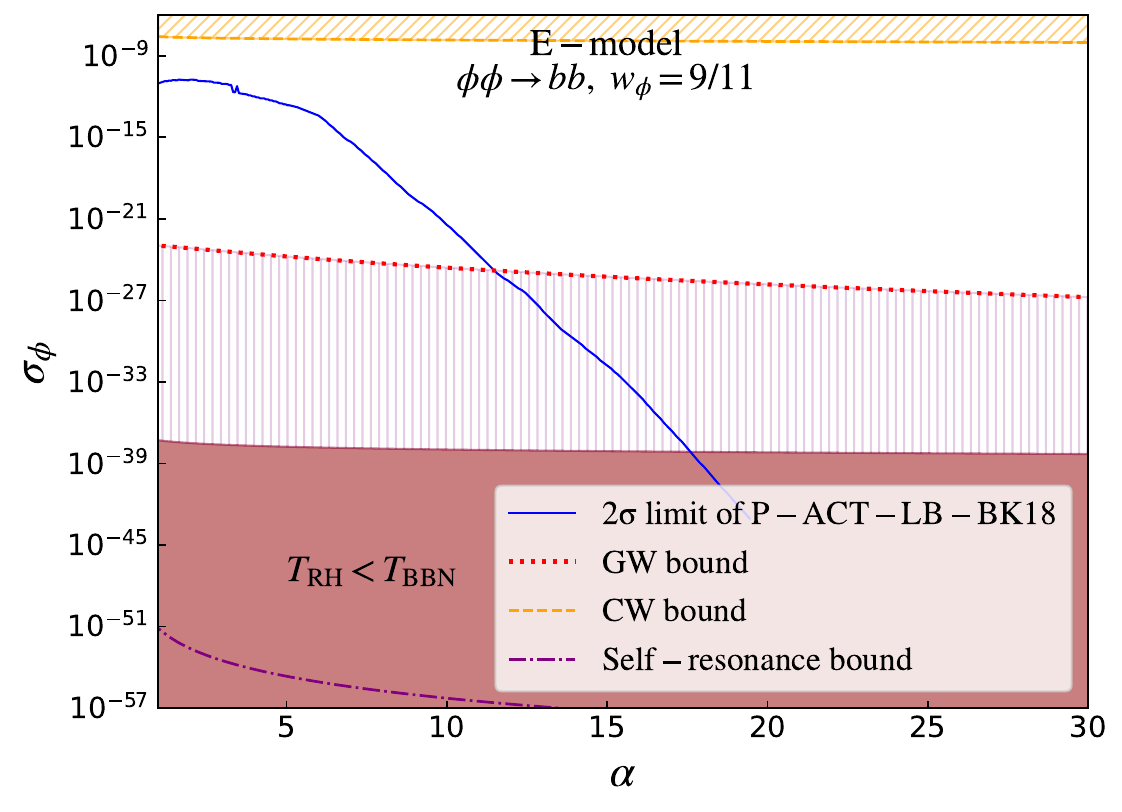}
    \caption[]{\justifying\it Constraints on the three different inflaton--SM couplings, namely $\phi \to f\bar{f}$ (upper panel), $\phi \to bb$ (middle panel), and $\phi\phi \to bb$ (lower panel), are shown as a function of the potential parameter $\alpha$ for the \textbf{E-model} of inflation. The blue line indicates the upper bound on the coupling parameter derived from the $2\sigma$ limits of the recent {\rm P-ACT-LB-BK18} data. In addition, we show the constraints from inflaton self-resonance (violet dot-dashed), CW one-loop corrections (orange dashed) and PGW overproduction (red dotted), each represented with different color patterns.}
    \label{fig:emodel_refinement}
\end{figure*}

\subsection{Model Constraints}
\label{subsec:model_constraint}

\subsubsection*{\textit{E-model}}

Figs.~\ref{fig:nsrE} and \ref{fig:w13} illustrate the predictions of the E-model in the $(n_s, r)$ plane for various combinations of the parameter $\alpha$ and the reheating EoS parameter $w_\phi$. For reheating scenarios with $w_\phi < 1/3$, the combined current dataset P-ACT-LB-BK18 imposes a lower bound on the reheating temperature, while the upper bound is determined by the assumption of instantaneous reheating. Conversely, for $w_\phi > 1/3$, the situation is reversed. The upper bound on the reheating temperature is set by the aforementioned dataset, whereas the lower bound is fixed by BBN or limits on the effective number of relativistic species $\Delta N_{\rm eff}$ considering the PGWs.
For the matter-like reheating case ($w_\phi = 0$), the model predictions lie marginally within the $2\sigma$ confidence region for certain values of $\alpha$. However, the well-known Higgs-Starobinsky model, corresponding to $\alpha = 1$ and $n = 1$ of $\alpha-$ attractor E-model, is excluded at the $2\sigma$ level based on the P-ACT-LB-BK18 dataset, as shown in the upper left panel of Fig. \ref{fig:nsrE}~\footnote{However, in Ref.~\cite{Haque:2025uis}, we have shown that by treating the reheating EoS as a free parameter, the Higgs-Starobinsky model remains consistent with the current observational data within the $2\sigma$ confidence level, even for a stiff EoS.}. As $w_\phi$ increases, the predicted values of $(n_s, r)$ shift rightward in the plane, moving into better agreement with observational constraints. This leads to a broader allowed parameter space, not only for inflationary parameters such as $\alpha$, but also for post-inflationary quantities like $T_{\text{RH}}$ and the inflaton coupling. 
However, for stiff EoS with $w_\phi \gtrsim 0.6$, the constraints from PGW overproduction become significant. For example, the upper limit on $\alpha$ for $w_\phi = 9/11$ is tightened from $16.5$ to $11.7$ due to the overproduction of PGWs. The detailed numerical results are presented in Table~\ref{tab:model_constraint_self_resonance} for E-model considering different values of $w_{\phi}$.
Our finding shows that $\w=1/3$ aligns with the $2\sigma$ limit of the dataset for $\alpha\in[4,7]$.

\begin{figure*}
    \centering
    \includegraphics[scale=0.3]{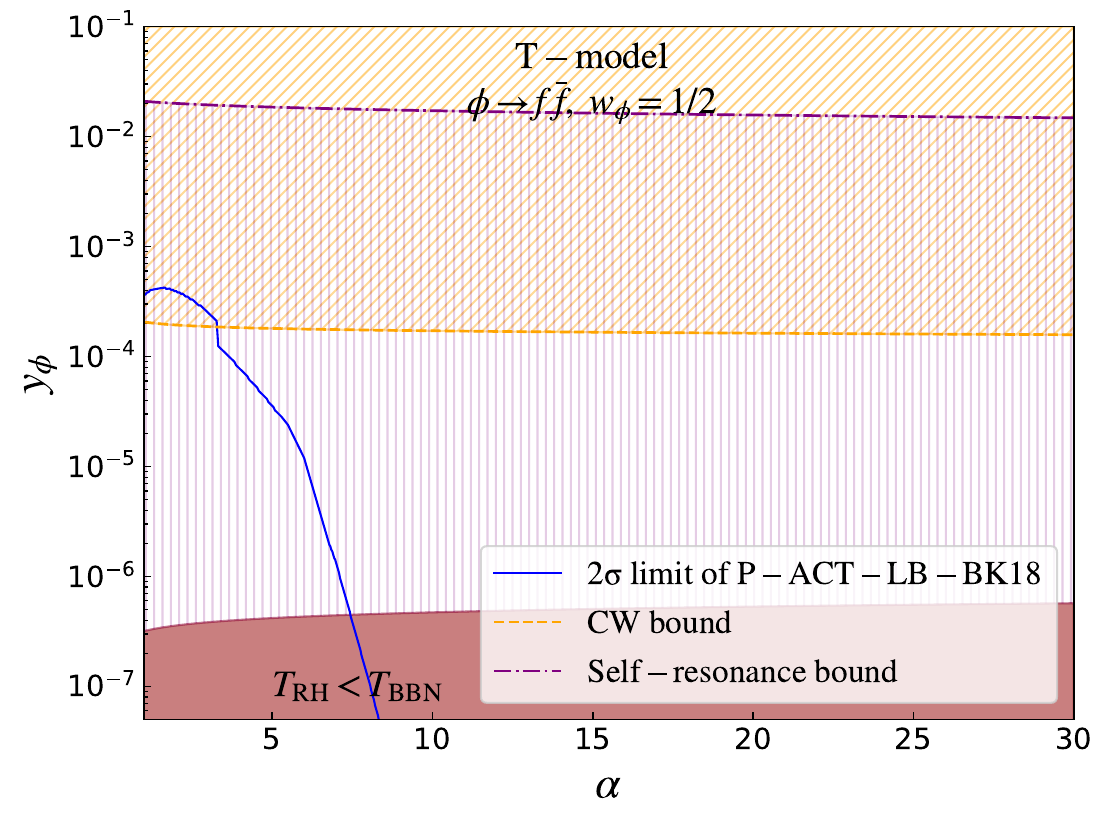}
    \includegraphics[scale=0.3]{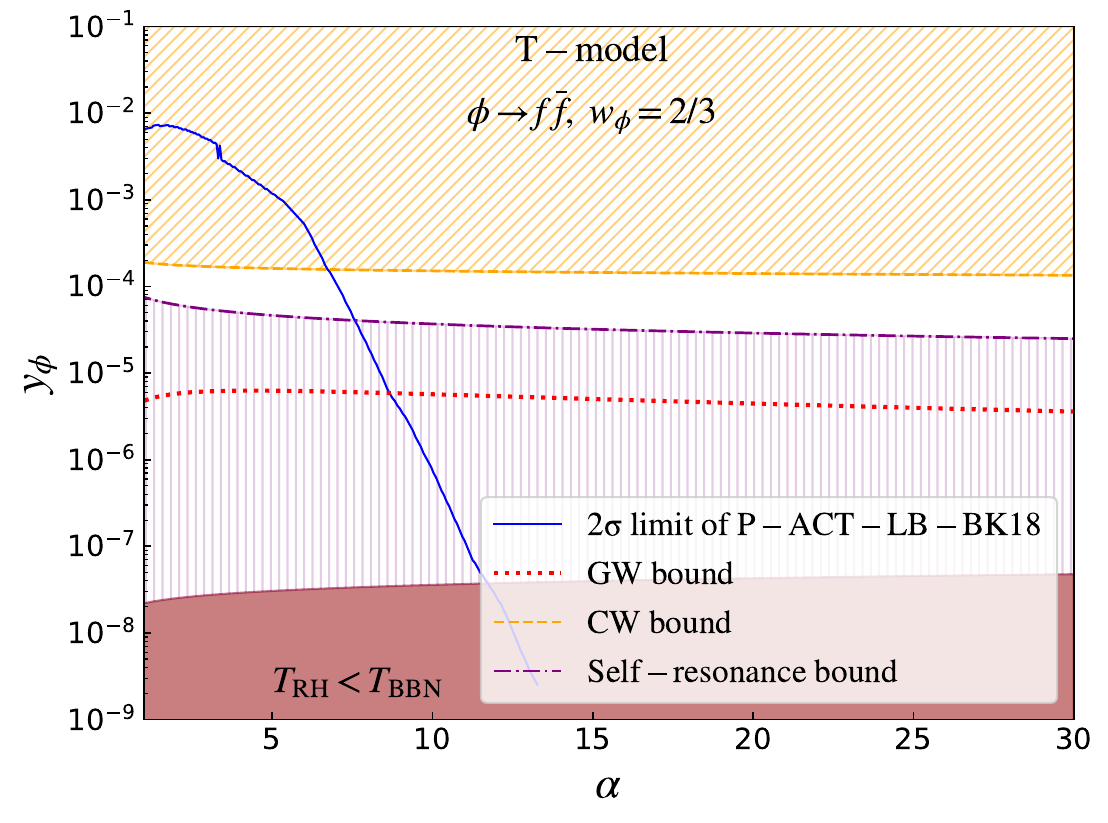}
    \includegraphics[scale=0.3]{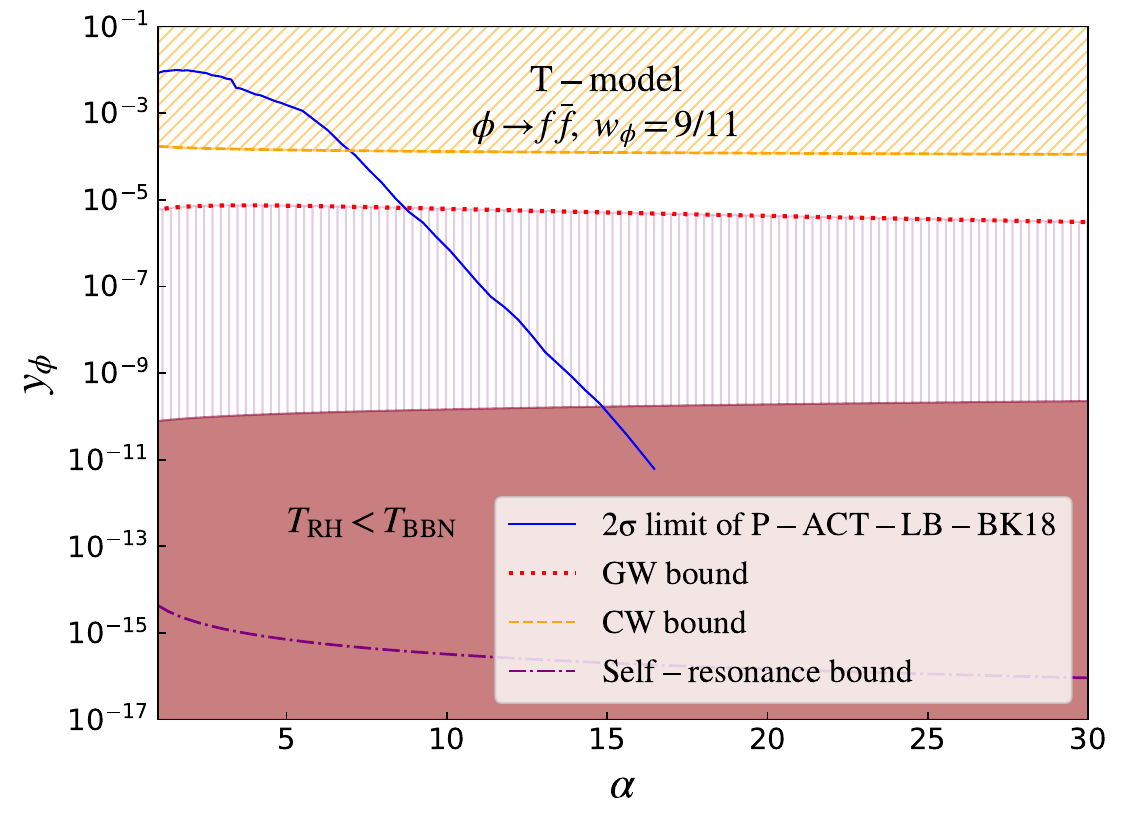}
    \includegraphics[scale=0.3]{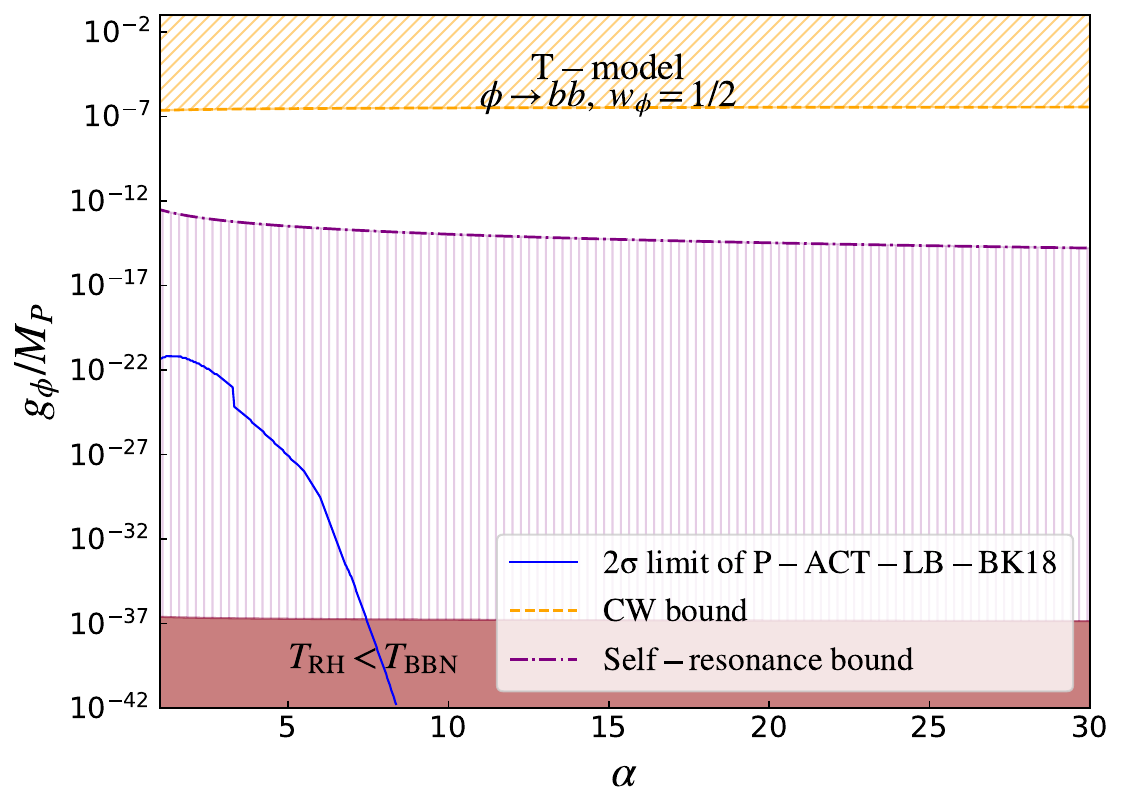}
    \includegraphics[scale=0.3]{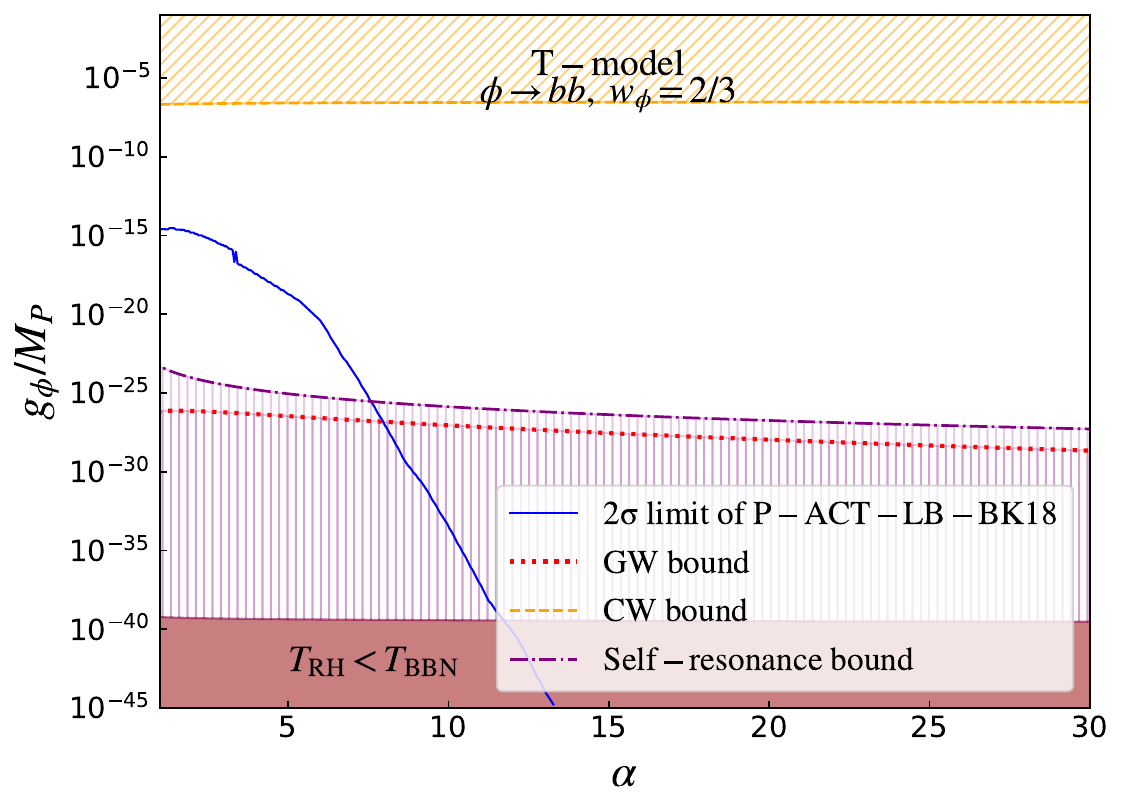}
    \includegraphics[scale=0.3]{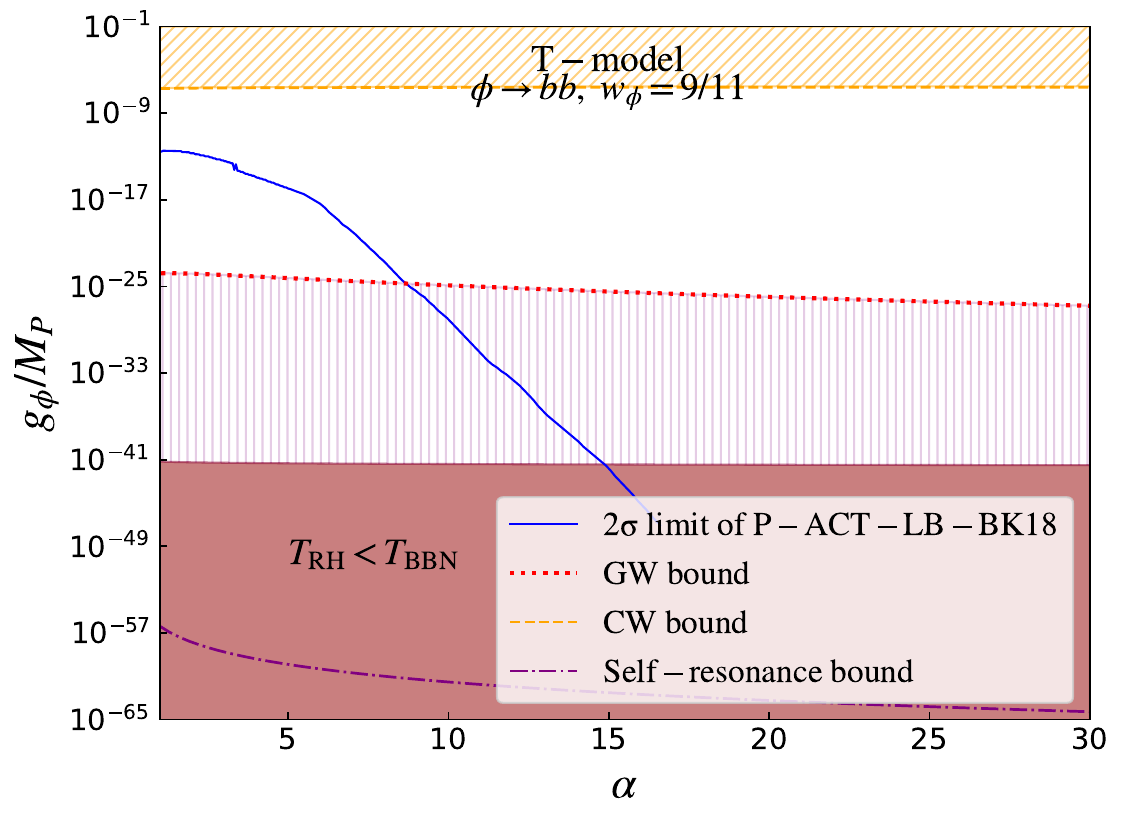}
    \includegraphics[scale=0.3]{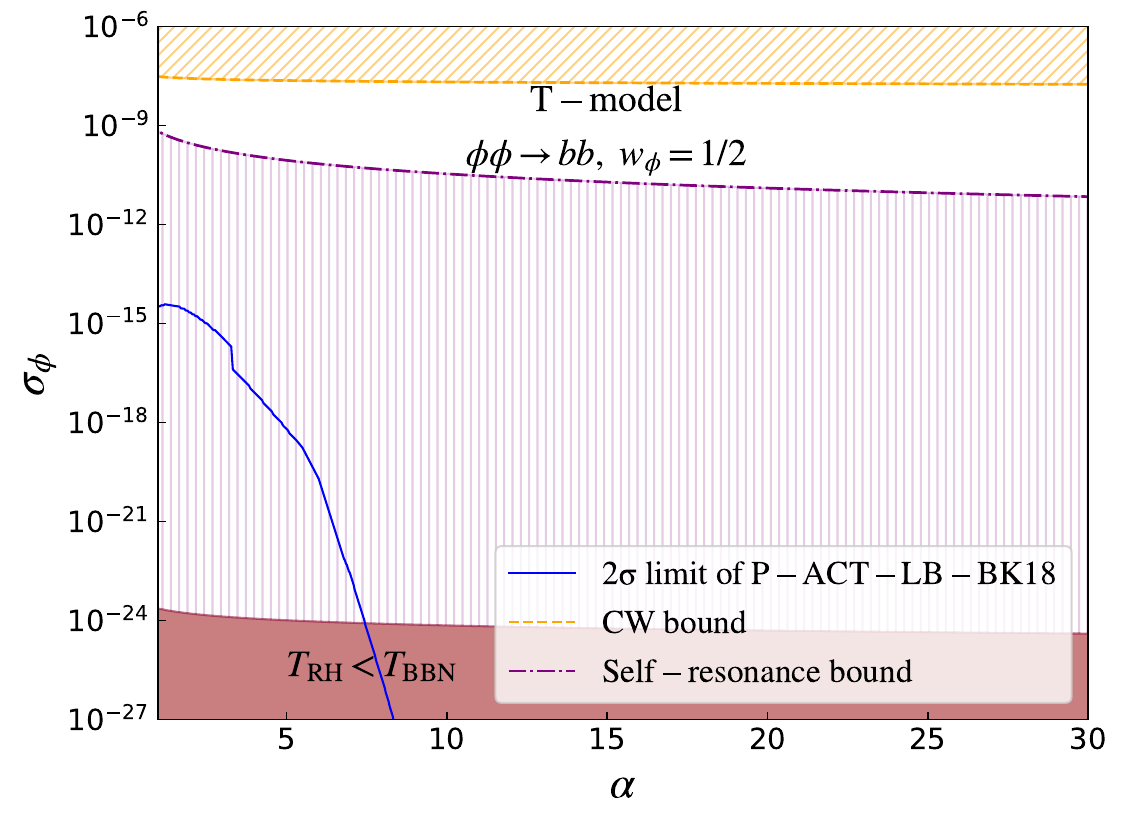}
    \includegraphics[scale=0.3]{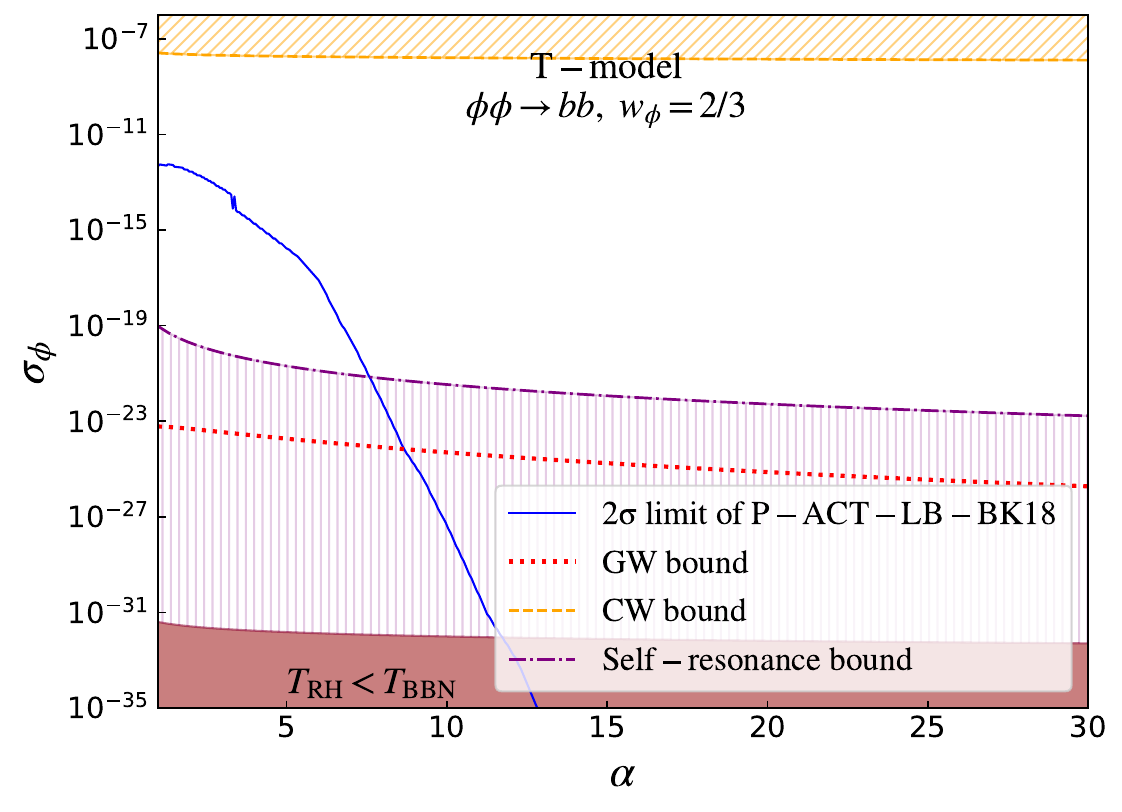}
    \includegraphics[scale=0.3]{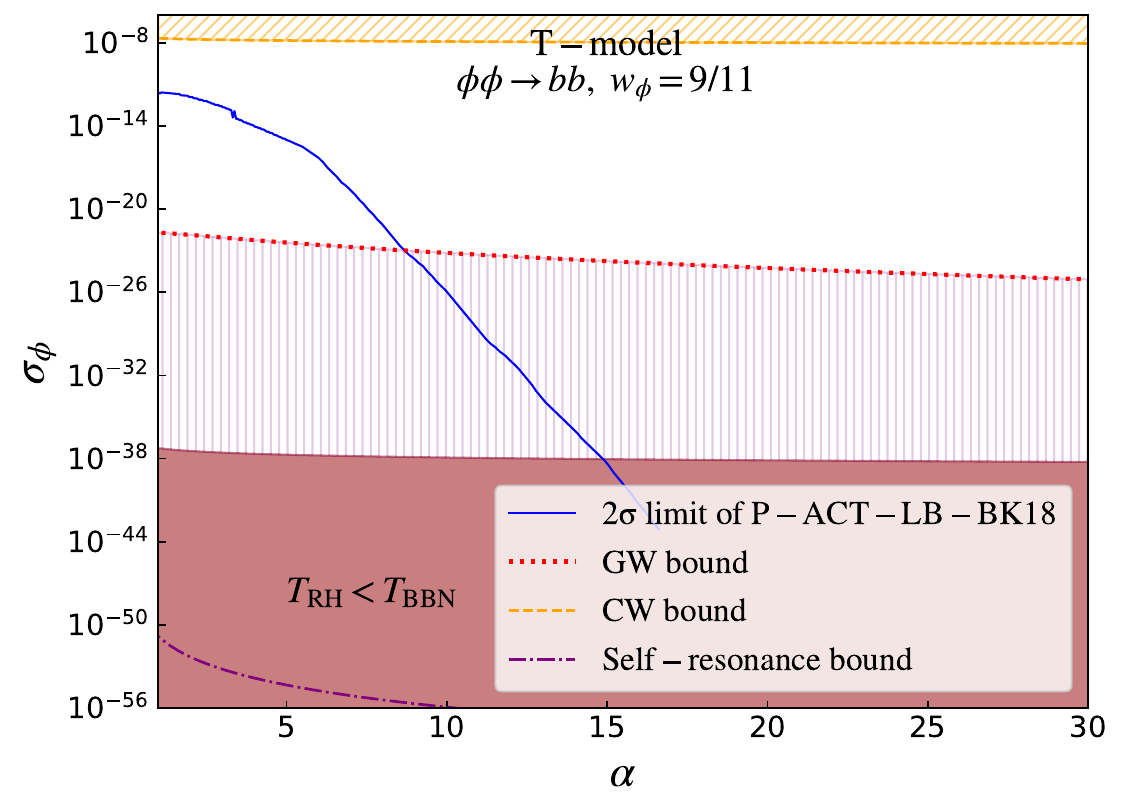}
    \caption[]{\justifying\it Description of the figure is same as Fig.~\ref{fig:emodel_refinement} but for the \textbf{T-model} of inflation.}
    \label{fig:tmodel_refinement}
\end{figure*}

We now examine how the $2\sigma$-consistent region in the $T_{\rm RH}$-$\alpha$ plane is modified once the constraints from both self-resonance and PGWs overproduction are taken into account, as shown in Fig.~\ref{fig:trh_self_resonance} for both the E- and T-model realizations of $\alpha$-attractor inflation. The self-resonance bound becomes relevant for $w_\phi \geq 1/3$, whereas the constraint from PGW overproduction is typically relevant for stiffer EoS, $w_\phi \gtrsim 0.6$ (see, for instance, Fig.~\ref{fig:trh_self_resonance}, where the PGW constraint is not relevant for $w_\phi = 1/2$). Once $w_\phi \geq 1/3$, the self-resonance constraint becomes operative and can dominate over both the PGW and BBN bounds on $T_{\rm RH}$, thereby yielding a stronger lower limit on the reheating temperature within which our perturbative analysis remains valid. For example, in the E-model with $w_\phi = 1/2$, the inclusion of the self-resonance condition significantly tightens the allowed reheating window, restricting it to a narrow range $T_{\rm RH} \in [4.7\times10^{9},\,7.9\times10^{10}]~\mathrm{GeV}$, as presented in Table~\ref{tab:model_constraint_self_resonance}.  Furthermore, as the parameter $n$ increases, the self-resonance driven reheating duration $N_{\rm RH}^{\rm (sr)}$ also increases (see Eq.~\eqref{eq:Nrh_selfresonance}), which relaxes the lower bound on the reheating temperature arising from self-resonance. Consequently, for sufficiently large $n$ or larger values of $w_\phi$, the PGW constraint can again become dominant.

\begin{table*}[!ht]
    \centering
    \renewcommand{\arraystretch}{1.2}
    \begin{tabular}{c|c|c|c|c|c|c|c|c}
    \hline
    \hline
        \multirow{2}{*}{$n(\w)$} & \multicolumn{2}{|c}{range of $\alpha$} & \multicolumn{2}{|c}{range of $N_{\rm k}$} & \multicolumn{2}{|c}{maximum $T_{\rm RH}$ [GeV]} & \multicolumn{2}{|c}{minimum $T_{\rm RH}$ [GeV]}\\
        \cline{2-9}
        & E-model & T-model & E-model & T-model & E-model & T-model & E-model & T-model\\
        \hline
        $1(0)$ & $[2.2-22.5]$ & $-$ & $[55.9-56.6]$ & $-$ & $2.52\times10^{15}$ & $-$ & $7.3\times10^{11}$ & $-$\\
        $3(1/2)$ & $[0.09-2.15]$ & $-$ & $[56.0-57.5]$ & $\textcolor{blue}-$ & $7.9\times10^{10}$ & $\textcolor{blue}-$ & $4.7\times10^{9}$ & $\textcolor{blue}-$\\
        $5(2/3)$ & $[0.06-11.5]$ & $[1-7.684]$ & $[56.5-62.4]$ & $[58.4-62.2]$ & $2.9\times10^{11}$ & $5.4\times10^{9}$ & $1.87\times10^3$ & $3.4\times10^{3}$\\
        $10(9/11)$ & $[0.06-9.7]$ & $[1-9.7]$ & $[56.9-61.6]$ & $[58.4-63.7]$ & $5.4\times10^{11}$ & $1.3\times10^{11}$ & $9.6\times10^{3}$ & $5.3\times10^{4}$\\
    \hline
    \hline
    \end{tabular}
    \caption[]{\justifying \it Limiting values of the inflationary model parameters ($\alpha, N_k$) and bounds on the reheating temperature, obtained using the $2\sigma$ limits from recent {\rm P-ACT-LB-BK18} data, consistent with self-resonance constraints and gravitational wave overproduction limits, for both the E- and T-models.}
    \label{tab:model_constraint_self_resonance}
\end{table*}

\subsubsection*{\textit{T-model}}

Figs.~\ref{fig:nsrT} and \ref{fig:w13} present the corresponding results for the T-model. A key distinction from the E-model is that, the case with $w_\phi \leq 1/3$ is {\it{entirely excluded}} at the $2\sigma$ level for any values of $\alpha$. The T-model enters the $2\sigma$ confidence region {\it only} for $w_\phi \gtrsim 0.44$, with the best fit occurring around $\alpha \sim 4$. This indicates a strong preference for a relatively stiff post-inflationary EoS within this class of models.
Similar to the E-model analysis, the inclusion of both self-resonance and PGW overproduction constraints further tightens the allowed $2\sigma$ region in the $T_{\rm RH}$--$\alpha$ plane, as shown in the right panel of Fig.~\ref{fig:trh_self_resonance}. Since the T-model generally predicts lower reheating temperatures than the E-model for the same value of $\alpha$, the self-resonance condition becomes more restrictive, significantly narrowing the allowed reheating window. In particular, we find that the self-resonance constraint \textit{completely excludes} the case $w_\phi = 1/2$ for all values of $\alpha$. Consequently, the allowed parameter space for $\alpha$, $N_k$, and $T_{\rm RH}$ is more constrained in the T-model compared to the E-model. Numerical bounds on $\alpha$, $N_k$, and $T_{\text{RH}}$, incorporating both GW-overproduction and self-resonance constraints for different $w_\phi$ are tabulated in Table~\ref{tab:model_constraint_self_resonance}, for both E- and T-model of inflation.

\begin{table*}[!ht]
    \centering
    \renewcommand{\arraystretch}{1.2}
    \begin{tabular}{c|c|c|c}
    \hline
    \hline
        \multirow{1}{*}{$n(\w)$} & \multicolumn{1}{|c}{range of $y_{\phi}$} & \multicolumn{1}{|c}{range of $g_{\phi}/\Mpl$} & \multicolumn{1}{|c}{range of $\sigma_{\phi}$}\\
        \hline
    $1(0)$ & $-$ &   $[2.1\times10^{-8}-5\times10^{-7}]$ & $-$\\
        $3(1/2)$ & $-$ &  $[7.3 \times 10^{-16}-1.1\times10^{-12}]$ &  $[2.8\times10^{-11}-6.3\times10^{-10}]$ \\
        $5(2/3)$ & $[2.5\times10^{-5}-2.1\times10^{-4}]$ & $[2.2\times10^{-28}-3.8\times10^{-12}]$ & $[1.1\times10^{-22}-8.7\times10^{-11}]$ \\
        $10(9/11)$ & $[1.2\times10^{-6}-1.8\times10^{-4}]$ & $[2.3\times10^{-29}-3.6\times10^{-12}]$ &  $[1.5\times10^{-25}-1.8\times10^{-11}]$\\
    \hline
    \hline
    \end{tabular}
    \caption[]{\justifying \it Bounds on different inflaton-SM couplings obtained using the $2\sigma$ limits from recent {\rm P-ACT-LB-BK18} data, consistent with self-resonance constraints and gravitational wave overproduction limits for \textbf{E-models}.}
    \label{tab:coupling_constraint_E_CW}
\end{table*}

\begin{table*}[!ht]
    \centering
    \renewcommand{\arraystretch}{1.2}
    \begin{tabular}{c|c|c|c}
    \hline
    \hline
        \multirow{1}{*}{$n(\w)$} & \multicolumn{1}{|c}{range of $y_{\phi}$} & \multicolumn{1}{|c}{range of $g_{\phi}/\Mpl$} & \multicolumn{1}{|c}{range of $\sigma_{\phi}$}\\
        \hline
        $1(0)$ & $-$ &  $-$ & $-$\\
        $3(1/2)$ & $-$ &  $-$ &  $-$ \\
        $5(2/3)$ & $[2.5\times10^{-5}-1.9\times10^{-4}]$ & $[5.2\times 10^{-28}-2\times10^{-15}]$ & $[9.8\times10^{-22}-2\times10^{-13}]$ \\
        $10(9/11)$ & $[3.1\times10^{-6}-1.7\times10^{-4}]$ & $[1.6\times10^{-27}-5.6\times10^{-13}]$ &  $[1\times10^{-23}-2.5\times10^{-12}]$\\
    \hline
    \hline
    \end{tabular}
    \caption[]{\justifying \it Description of table is same as Table~\ref{tab:coupling_constraint_E_CW} but for \textbf{T-models}.}
    \label{tab:coupling_constraint_T_CW}
\end{table*}

\subsection{Constraints on Coupling Parameters}
\label{subsec:coupling_parameter}
Inflaton--SM couplings, whether through bosonic or fermionic decay channels, as well as through scattering processes, are directly tied to the reheating temperature and the number of e-foldings, and are thus constrained by the aforementioned combined observations. In addition to the constraints discussed in the previous subsection-namely, the restriction from the $2\sigma$-consistent region allowed by {\rm P-ACT-LB-BK18} data, together with bounds from self-resonance and PGW overproduction, we also incorporate the CW one-loop radiative correction to the inflationary potential. These corrections cannot be neglected and become particularly relevant in certain regions of parameter space.  Figures~\ref{fig:emodel_refinement_w0}-\ref{fig:tmodel_refinement} show the $2\sigma$-allowed ranges in the coupling strength, upon which we impose the constraints from PGW overproduction, the CW bound, and self-resonance. The resulting limits are presented for the Yukawa coupling $y_\phi$, the bosonic coupling $g_\phi$, and $\sigma_\phi$ in both E- and T-models.

Fig.~\ref{fig:emodel_refinement_w0} presents the results for the E-model of inflation with $w_\phi = 0$ scenario, considering the decay channels $\phi \to f\bar{f}$ (left panel) and $\phi \to bb$ (right panel). For the inflaton scattering process $\phi\phi \to bb$, the $w_\phi = 0$ case is not viable, as reheating cannot be successfully achieved. We do not show results for the T-model here, since for the T-model with $n=1$ (corresponding to $w_\phi = 0$), the predictions are inconsistent with the latest P-ACT-LB-BK18 data (see, for example, the upper left panel of Fig.~\ref{fig:nsrT}). On the other hand, constraints on the couplings for different values of $w_\phi = (1/2,\,2/3,\,9/11)$ are presented in Fig.~\ref{fig:emodel_refinement} and Fig.~\ref{fig:tmodel_refinement} for the E- and T-models, respectively. Comparing Figs.~\ref{fig:emodel_refinement_w0} and \ref{fig:emodel_refinement}, we find that for the E-model, the Yukawa coupling $y_\phi$ is \textit{completely excluded} at the $2\sigma$ level for $w_\phi \leq 1/2$, for all values of $\alpha$. For larger values of $w_\phi$, strong coupling regimes are constrained by the CW upper bound, while the small-coupling region is excluded by the combined PGW bound and self-resonance constraints. Consequently, only an intermediate range of couplings remains viable at the $2\sigma$ level, consistent with the constraints from self-resonance and PGW overproduction. For instance, for $w_\phi = 2/3$, we find $y_\phi \in [10^{-5},\,10^{-4}]$ to be allowed for both the E- and T-models. Note that as $w_\phi$ increases towards a very stiff EoS ($w_\phi > 2/3$), the self-resonance constraint becomes less efficient, and the allowed Yukawa coupling range widens. For example, for $w_\phi = 9/11$, the allowed Yukawa coupling consistent with observational data and the above constraints becomes $y_\phi \in (10^{-6},\,10^{-4})$ for both the E- and T-models of attractor inflation.

In contrast to the fermionic decay channel, the bosonic decay scenario, $\phi \to bb$, admits a comparatively wider viable $2\sigma$ parameter space. For instance, when $w_\phi = 0$, as shown in the right panel of Fig.~\ref{fig:emodel_refinement_w0}, the upper bound on $g_{\phi}$ arises solely from the CW radiative correction, yielding a narrow viable region; for the E-model, this corresponds to $g_{\phi}/M_{\rm P} \in (10^{-8},\,10^{-6})$. However, for larger values of $w_\phi$, the maximum value of $g_{\phi}$ allowed by the $2\sigma$ P-ACT-LB-BK18 constraint lies well below the CW limit, indicating that perturbative reheating scenarios consistent with observational data are effectively insensitive to CW radiative corrections (\textit{cf.}\ middle panels of Figs.~9 and 10 for the E- and T-models, respectively). In these cases, \textit{i.e.}, $w_\phi \geq 1/2$, the upper bound on the coupling strength is set by the $2\sigma$ P-ACT-LB-BK18 constraint, while the lower bound on $g_{\phi}$ is determined by self-resonance and PGW constraints: for $w_\phi = 1/2$ and $2/3$, the self-resonance bound dominates, whereas for very stiff EoS (\textit{e.g.}, $w_\phi = 9/11$), the PGW overproduction constraint becomes more stringent. This conclusion holds for both the E- and T-models of inflation; however, for the T-model with $w_\phi = 1/2$, no parameter space remains consistent with observations once the self-resonance constraint is imposed. For another process, the inflaton-scattering channel, $\phi\phi \to bb$, the case $w_\phi = 0$ is not viable since reheating cannot be achieved. Nevertheless, a viable range of the scattering parameter $\sigma_{\phi}$ re-emerges for sufficiently stiff EoS, $w_\phi \geq 1/2$ (see, \textit{e.g.}, the lower panels of Figs.~\ref{fig:emodel_refinement} and \ref{fig:tmodel_refinement}), where the combined PGW and self-resonance constraints set the lower bound on $\sigma_{\phi}$, similar to the bosonic decay channel $\phi \to bb$. In this bosonic scattering scenario for stiff EoS, the CW bound is not effective and does not alter the conclusions regarding the allowed $2\sigma$ region. All results are summarised in Tables~\ref{tab:coupling_constraint_E_CW} and \ref{tab:coupling_constraint_T_CW} and illustrated in the Figs.~\ref{fig:emodel_refinement_w0}, ~\ref{fig:emodel_refinement} and \ref{fig:tmodel_refinement}.

\section{Conclusions}
\label{sec:conclusion}
In this paper, we have studied the impact of recent {\rm P-ACT-LB-BK18} observations on $\alpha$-attractor inflationary potentials, focusing on both the E- and T-models of inflation as well as the post-inflationary reheating phase, considering a perturbative reheating scenario. In addition to the BBN lower bound on the reheating temperature, we have incorporated constraints from primordial gravitational wave overproduction, inferred from limits on the effective number of relativistic species, $\Delta N_{\rm eff}$. Beyond these observational constraints, we also consider the effects of inflaton self-resonance and CW one-loop radiative corrections to the inflationary potential, which can, in principle, modify our results and impose additional theoretical constraints, thereby reducing the $2\sigma$-allowed parameter space consistent with latest data. In particular, larger values of the inflaton-SM couplings enhance CW corrections, leading to an upper bound on the coupling parameters, beyond which the CW-corrected potential must be taken into account and our perturbative  reheating analysis with the tree-level potential ceases to be valid. On the other hand, self-resonance can reheat the Universe through non-linear dynamics even in the absence of direct couplings to SM fields, resulting in non-perturbative particle production. We estimate that if the number of reheating e-folds associated with self-resonance satisfies $N_{\rm RH}^{(\rm sr)} < N_{\rm RH}$, then the perturbative reheating scenario is no longer reliable. Within the coupling ranges allowed by our combined observational and theoretical constraints, our analysis not only provides updated bounds on the parameters governing these inflationary potentials but also extends to the post-inflationary reheating phase. In particular, we constrain the reheating temperature and translate these bounds into limits on three representative inflaton-SM interaction channels: (a) $\phi \to \bar{f}f$, (b) $\phi \to bb$, and (c) $\phi\phi \to bb$. In the following, we summarize our key findings.

\begin{itemize}
    \item For the E-model, we find that reheating scenarios with an inflaton EoS $\w \leq 1/3$ are allowed within the $2\sigma$ C.L.\ for specific ranges of the $\alpha$ parameter. In particular, matter-like reheating ($\w = 0$) is consistent with current data for $\alpha \in [2.2,\,22.5]$. In contrast, the T-model exhibits significantly tighter constraints: the entire regime with $\w \leq 1/3$ is excluded at the $2\sigma$ C.L. Although for $w_\phi = 1/2$ both the E- and T-models admit an allowed region at the $2\sigma$ level, this region lies in the self-resonance dominated regime for T- model, where the assumption of perturbative reheating is no longer valid, and hence these results cannot be reliably trusted within our framework. 
    \item In order to constrain the post-inflationary dynamics consistent with the latest {\rm P-ACT-LB-BK18} data, we find that for the E-model with $n=1$ ($w_\phi = 0$), the maximum allowed reheating temperature is approximately $T_{\rm RH} \sim 3 \times 10^{15}\,\mathrm{GeV}$, determined by the maximum allowed Hubble parameter at the end of inflation, $H_{\rm end}$, assuming instantaneous reheating. This upper bound is roughhly same for $w_\phi \leq 1/3$. The lower bound on the reheating temperature is set by consistency with observational data; for $w_\phi = 0$, we find $T_{\rm RH} \sim 7 \times 10^{11}\,\mathrm{GeV}$. In contrast, the T-model of inflation is not consistent with the latest $2\sigma$ data over the full range of reheating temperature for $w_\phi \leq 1/3$. However, as the EoS parameter increases ($w_\phi > 1/3$), the upper bound on the reheating temperature is set by the {\rm P-ACT-LB-BK18} constraints. Importantly, we incorporate limits from the primordial gravitational wave background, which become particularly relevant for stiff EoS ($w_\phi \gtrsim 0.6$), effectively raising the lower bound on the reheating temperature beyond the standard BBN requirement. In addition, we include constraints from self-resonance, which restrict the region of smaller reheating temperatures where the Universe reheats through non-linear dynamics, rendering the perturbative reheating framework invalid. For $1/3 \lesssim w_\phi \lesssim 0.7$, the self-resonance constraint provides a dominant contribution compared to the PGW bound, thereby imposing a stringent lower limit on the reheating temperature. Comprehensive numerical results are summarized in Table~\ref{tab:model_constraint_self_resonance}. Note that in Table~\ref{tab:model_constraint_self_resonance}, where we present constraints on the inflationary and post-inflationary parameters, we do not include the CW  one-loop radiative correction constraints, since no specific coupling values are assumed. As the CW bound depends explicitly on the coupling strength, it must be taken into account separately when assessing the consistency of the inflaton--SM couplings with observational data.

    \item In the perturbative reheating setup, different inflaton decay and scattering processes, namely, the three non-gravitational interaction channels of interest in Eq.~\ref{eq:Lint}, determine the reheating temperature and thus directly connect the inflaton-SM couplings to observational and theoretical constraints. Our results show that, in the E-model, the Yukawa coupling $y_\phi$ is completely excluded at the $2\sigma$ level for $w_\phi \leq 1/2$ for all values of $\alpha$; a similar conclusion holds for the T-model. For stiffer EoS, only an intermediate window remains viable, bounded from above by CW  radiative corrections and from below by the combined PGW and self-resonance constraints. Similar to the Yukawa case, the bosonic decay coupling $g_\phi$ and the bosonic scattering strength $\sigma_\phi$ are excluded for $w_\phi \leq 1/2$ in the T-model. However, in the E-model, viable regions remain, as summarized in Table~\ref{tab:coupling_constraint_E_CW}. In particular, for the E-model, the bosonic decay channel admits viable parameter space for all values of the EoS. Overall, we find that the T-model is more tightly constrained compared to the E-model: all couplings are excluded at $2\sigma$ for $w_\phi \leq 1/2$, and even for stiffer EoS only a narrow intermediate range survives, with CW corrections limiting large couplings and the combined PGW and self-resonance bounds setting the dominant lower limits. Detailed numerical ranges for all couplings in both models are summarized in Tables~\ref{tab:coupling_constraint_E_CW} and \ref{tab:coupling_constraint_T_CW}.
\end{itemize}

Looking ahead, forthcoming cosmological observations offer strong prospects for advancing our understanding of the early Universe and physics beyond the SM. In particular, next-generation experiments such as CMB-S4~\cite{CMB-S4:2020lpa} and LiteBIRD~\cite{LiteBIRD:2022cnt} are expected to significantly tighten constraints in the $(n_s, r)$ parameter space. These improved measurements will enhance the precision of the primordial power spectrum and inflationary dynamics, helping to distinguish between competing inflationary models. Consequently, they will provide deeper insights into the nature of the inflaton and its interactions, potentially revealing new aspects of high-energy physics. Overall, the synergy between theoretical developments and high-precision observations will remain crucial, with upcoming experiments poised to play a transformative role in this effort.\\

\section*{Acknowledgments}
 We thank the anonymous referee for their valuable comments and suggestions, which have greatly improved the quality of this work. MRH acknowledges the Tsung-Dao Lee Institute, Shanghai Jiao Tong University, for financial support through the Siyuan
Postdoctoral Fellowship.  SP thanks CSIR for financial support through Senior Research Fellowship (File no. 09/093(0195)/2020-EMR-I). DP thanks ISI Kolkata for financial support through Senior Research Fellowship.


\bibliographystyle{apsrev4-2}
\bibliography{references}
\end{document}